 \documentclass[preprint]{aastex6}

\usepackage{natbib}
\usepackage{amsmath}
\usepackage{multirow}
\usepackage{color,colortbl}

\newcommand{\myemail}{billylquarles@gmail.com}
\newcommand{\aCenA}{$\alpha$ Cen A}
\newcommand{\aCenB}{$\alpha$ Cen B}
\definecolor{Gray}{gray}{0.85}

\shorttitle{Planet Packing in $\alpha$ Cen AB}
\shortauthors{Quarles \& Lissauer}

\begin{document}


\title{Long-Term Stability of Tightly Packed Multi-Planet Systems in Prograde, Coplanar, Circumstellar Orbits within the $\alpha$ Centauri AB System}


\author{B. Quarles\altaffilmark{1}}
\affil{HL Dodge Department of Physics \& Astronomy, University of Oklahoma, Norman, OK 73019, USA}
\affil{NASA Ames Research Center,  Space Science and Astrobiology Division, MS 245-3,
    Moffett Field, CA 94035}
\email{\myemail}
\author{Jack J. Lissauer}
\affil{NASA Ames Research Center, Space Science and Astrobiology Division, MS 245-3,
    Moffett Field, CA 94035}


\altaffiltext{1}{NASA Postdoctoral Fellow}


\begin{abstract}
We perform  long-term simulations, up to ten billion years, of closely-spaced configurations of 2 -- 6 planets, each as massive as the Earth, traveling on nested orbits about either stellar component in $\alpha$ Centauri AB.  The innermost planet initially orbits at either the inner edge of its star's empirical habitable zone (HZ) or the inner edge of its star's conservative HZ.  Although individual planets on low inclination, low eccentricity, orbits can survive throughout the habitable zones of both stars, perturbations from the companion star require that the minimum spacing of planets in multi-planet systems within the habitable zones of each star must be significantly larger than the spacing of similar multi-planet systems orbiting single stars in order to be long-lived.  The binary companion induces a forced eccentricity upon the orbits of planets in orbit around either star. Planets on appropriately-phased circumstellar orbits with initial eccentricities equal to their forced eccentricities can survive on more closely spaced orbits than those with initially circular orbits, although the required spacing remains higher than for planets orbiting single stars.  A total of up to nine planets on nested prograde orbits can survive for the current age of the system within the empirical HZs of the two stars, with five of these orbiting  $\alpha$ Centauri B and four orbiting  $\alpha$ Centauri A.
\end{abstract}


\keywords{}



\section{Introduction}
The $\alpha$ Centauri  system contains three stars. The two largest stars in the system, \aCenA{} and \aCenB, travel about one another on an eccentric orbit with a periapsis of $\sim$~11 AU and are broadly similar to the Sun in mass (1.133 M$_\odot$ and 0.972 M$_\odot$, \cite{Pourbaix2016}) and luminosity (1.519 L$_\odot$ and 0.5 L$_\odot$, \cite{Thevenin2002}). The loosely-bound third star, $\alpha$ Centauri C, is much smaller and fainter and has an estimated orbital semimajor axis $\sim$~9,000 AU from the center of mass \citep{Kervella2016b}; $\alpha$ Cen C is usually referred to as Proxima Centauri because it is the nearest star to our Solar System. Models of planetary accretion suggest that circumstellar planets could have formed around \aCenA{} and \aCenB{} \citep{Quintana2002,Quintana2003}, provided the collision velocities of late stage planetesimals are not too large \citep{Thebault2008,Thebault2009,Thebault2014}.  Recent studies have investigated the necessary disk conditions that produce favorable environments for accretion to dominate over fragmentation, leading to formation of systems of multiple planets \citep{Xie2008,Xie2010,Rafikov2015a,Rafikov2015b,Silsbee2015}.  

Public and scientific interest in the $\alpha$ Centauri AB system has increased in the past few years due to the abundance of exoplanetary detections by the {\it Kepler} mission and the recent discovery of an Earth-mass planet orbiting Proxima Centauri, Proxima b \citep{Gillon2016}.  Additionally, recent observational studies have indicated the possible existence of a close-in planet $\alpha$ Cen B b \citep{Dumusque2012}, but scrutiny of the analysis methods have placed the claimed detection into doubt \citep{Hatzes2013,Rajpaul2016}.  Observations of the binary system continue with the possible observation (single event) of a transiting planet on a somewhat more distant orbit about $\alpha$ Cen B by \cite{Demory2015}, and a ground-based radial velocity campaign has ruled out the presence of very massive close-in planets \citep{Endl2015}.  {\cite{Zhao2017} have shown that existing data places mass detection thresholds for planets in the classically defined habitable zones (HZs) at $\lesssim 50$ M$_\oplus$ for $\alpha$ Cen A and $\lesssim 10$ M$_\oplus$ for $\alpha$ Cen B.}  Moreover, the possibility of a space mission that will directly image any planets that may orbit within the habitable zones (HZs) of $\alpha$ Cen A or $\alpha$ Cen B is currently being studied \citep{Belikov2015a,Belikov2015b,Bendek2015}.  Close conjunctions of the $\alpha$ Cen pair in the future may also provide a microlensing probe for the detection of planets in the system \citep{Kervella2016a}.  Our study investigates how the orbits of terrestrial planet systems would evolve on timescales commensurate with the estimated age of the system ($\sim$~5 -- 7 Gyr, according to \cite{Mamajek2008}).

\cite{Gladman1993} used Hill stability and numerical simulation to determine the minimum spacing of two Earth-mass planets orbiting a Sun-like star.  He found that an interplanetary spacing of semimajor axes of $\beta>2\sqrt{3}$ in units of the mutual Hill radius of the planets (Eq.~\ref{eqn:hill}) ensures stability of the system, whereas even slightly more closely-spaced planets suffer a close encounter very rapidly. \cite{Deck2013} showed that eccentric, coplanar systems behave in a more chaotic manner (due to the overlap of first-order MMRs) that depends on the weighted eccentricity and the mass ratio of the planets.  

{Systems containing four or more massive bodies are fundamentally different from the three body systems studied by \cite{Hill1878a,Hill1878b,Hill1878c}, \cite{Gladman1993}, and \cite{Deck2013} in that they have more degrees of freedom and are not well confined by integrals of motion.}  Studies of the planetary spacing required for stability of these many-body systems have relied primarily on numerical integrations. \cite{Chambers1996} further investigated the stability of multiple planetary systems considering up to twenty planets orbiting a Sun-like star, where the interplanetary spacing was normalized relative to the mutual Hill radius.  More recent works, \citep[e.g.,][]{Smith2009,Pu2015,Obertas2016} have further investigated different aspects of planetary packing involving the variations in the total number, masses, and resonant structure of planets.  

{Planetary packing in binary star systems has not been extensively explored, due in part to the high dimensionality of the parameter space and  the relatively few exoplanets discovered  in close binary systems like $\alpha$ Cen\footnote{\url{http://www.univie.ac.at/adg/schwarz/multiple.html}}.  Thus, we are motivated to engage in these studies using the techniques developed for single star systems.   We build upon our previous studies of orbital stability \citep{Quarles2016} and minimization of free eccentricity \citep{Quarles2017b}  of individual planets in the $\alpha$ Cenauri AB system and focus on the number of planets that could reside within each star's HZ.}

This paper presents the results of simulations of planetary systems orbiting either star in the $\alpha$ Cen AB system to identify how many 1 Earth mass (M$_\oplus$) planets could survive on nested, prograde, coplanar orbits within each of the stars' HZs.  Our methods are outlined in Section 2.  The results of our study are presented in Section 3.  We discuss the results of our study and compare them to analogous published studies of planet packing around single stars in Section 4.  In Section 5, we provide the general conclusions of our work along with the implications for future work.

\section{Methodology} \label{Method}
The numerical simulations in this work use a custom version of the {\tt mercury6} integration package that has been optimized for the study of planetary bodies within binary star systems \citep{Chambers2002}.  Using this numerical tool, we investigate the stability of tightly packed planets within the $\alpha$ Centauri AB binary star system on a ten billion year timescale.  We ignore the presence of $\alpha$ Cen C even though it resides in a loosely bound orbit because its pericenter distance, $\sim$~5,300 AU \citep{Kervella2016b}, is large compared to the sizes of both the AB binary orbit and the circumstellar habitable zones.  Recent studies by \cite{Worth2016} show that planetary systems near the larger stars can still form despite the possible turbulent evolution of the triple star system at earlier times.  We also ignore changes in the binary stars due to stellar evolution because we are largely sampling possible past states, where the dynamical effects due to stellar evolution are negligible.

\begin{figure*}
{\epsscale{1.}
\plotone{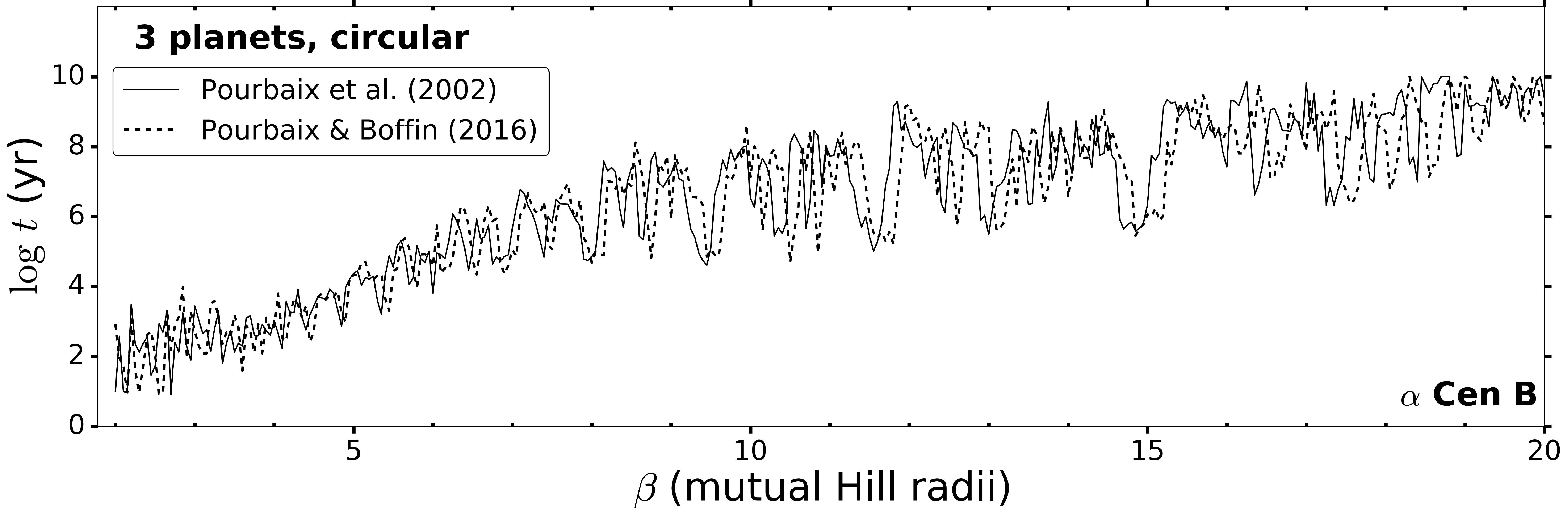}}
\caption{ Comparison of the lifetimes of three planet systems orbiting $\alpha$ Cen B using the older (solid curve) and the newer (dashed curve) parameters for the binary system.  The interplanetary spacings $\beta$ are measured in terms of the mutual Hill radius, which is slightly smaller for the newer parameters because the estimated mass of their host star has increased (see Eq.~\ref{eqn:iter}), thereby resulting in slightly larger values for $\beta$ for a given ratio of orbital periods of neighboring 1~M$_\oplus$ planets. The results of individual runs are connected with a solid or dashed line in order to emphasize the displacements caused by the differences in resonance locations for different stellar parameters. \label{bin_comp}}
\end{figure*}

In our study, we define long-term stability as the lack of collisions between and/or ejections of any of the planets that we numerically simulate up to the estimated age of the stellar binary system, 6 Gyr.  {A collision occurs when the distance between two bodies during an encounter is less than the sum of their radii.  A planet is considered ejected when its distance from the host star exceeds 100 AU.  Almost all of our simulations are stopped due to collisions.}  We keep the initial orbit of the binary stars fixed while varying the number and orbital spacing of a set of coplanar Earth-mass (1 M$_\oplus$) planets.  For the binary orbit, we adopt the revised orbital parameters from \cite{Pourbaix2016} and utilize the initial mean anomaly prescribed in \cite{Quarles2016}.  The dynamical outcomes do not change appreciably if one uses the parameters from \cite{Pourbaix2002}, as can be seen in Figure \ref{bin_comp}.  As all of the planets travel on prograde paths in the orbital plane of the binary stars, we adopt the dynamicist convention where $i_{bin} = 0^\circ$  denotes the reference plane and the longitude of pericenter is redefined as the argument of pericenter ($\varpi_{bin} \rightarrow \omega_{bin}$).

We select initial conditions for the terrestrial planetary systems following the method of \cite{Smith2009,Smith2010}, where the planets begin on coplanar orbits around the host star, $\alpha$ Cen A (1.133 M$_\odot$) or $\alpha$ Cen B (0.972 M$_\odot$).  The initial semimajor axes of these orbits vary based on an input spacing parameter $\beta$ and follow an iterative scheme \citep{Smith2009}.  In this procedure, the amount of mass that lies interior to the $j^{th}$ planet is given by: $\widetilde{M}_j = M_\star + (j-1) M_\oplus$.  We then use $\widetilde{M}_j$ in a recurrence relation to assign the semimajor axis, $a_{j+1}$, of the next planet. The value of $a_{j+1}$ depends upon the parameter $\beta$ and the mutual Hill radius for two adjacent planets, 

\begin{align} \label{eqn:hill}
R_{H_{j,j+1}} = {(a_j + a_{j+1}) \over 2}\left[{m_j + m_{j+1} \over 3\widetilde{M}_j}\right]^{1/3} 
\end{align}

\noindent \citep{Chambers1996}, as follows:
 
\begin{align} \label{eqn:iter}
a_{j+1} &= a_j + \beta R_{H_{j,j+1}} \nonumber \\
 &= a_j + {\beta \over 2} (a_j + a_{j+1})\left[{m_j + m_{j+1} \over 3\widetilde{M}_j}\right]^{1/3} \nonumber \\
 &=  a_j\left[1 + {\beta \over 2}\left({m_j + m_{j+1} \over 3\widetilde{M}_j}\right)^{1/3}\right]\left[1 - {\beta \over 2}\left({m_j + m_{j+1} \over 3\widetilde{M}_j}\right)^{1/3}\right]^{-1}.
\end{align}

\noindent This procedure yields systems where the initial period ratios between adjacent planets are roughly equal.  

The boundaries of the habitable zone around each star can be defined relative to the empirical limits derived for the Solar System or using more conservative limits that place the inner boundary so that it receives an amount of flux due to the radiation from its host star, $S$, equal to that intercepted by the Earth\footnote{The maximum amount of energy received by such a planet from the companion star is $\sim$1.3\% of that from the star that it orbits, which is intercepted by a planet orbiting $\alpha$ Cen B when the stars are at periastron and the planet lies along the line segment connecting the stellar positions.}, S$_\oplus$.  We employ both definitions as the actual conditions for planetary habitability are not generally well-defined and depend on several assumed parameters, such as atmospheric composition and evolution.  The inner edge of the empirical habitable zone (1.78 S$_\oplus$) is represented by $r_{\rm emp}$ and the outer edge (0.32 S$_\oplus$) by $r_{\rm out}$ \citep{Kasting1993,Kopparapu2013}.  Our definition of the conservative habitable zone places the inner boundary at $r_\oplus$ (1 S$_\oplus$) and uses the same outer edge as $r_{\rm out}$ of the empirical HZ.  The periastron distance between the stars is quite large compared to $r_{\rm out}$, so including irradiation by the stellar companion would only change the boundaries of the HZs by a small amount \citep{Eggl2012,Kaltenegger2013}, and thus we ignore heating by the companion star. The numerical values of the boundaries of the HZs relative to the host stars are given in Table \ref{tab:HZ}.  

\begin{deluxetable}{cccccccccc}
\tablecolumns{10}
\tablewidth{0pc}
\tablecaption{Boundary Limits Around Each Star \label{tab:HZ}}
\tablehead{\colhead{Star} & \colhead{Luminosity} & \colhead{$r_{\rm emp}$} & \colhead{$T_{i}$} & \colhead{$T_{bin}/T_{i}$} & \colhead{$r_\oplus$} & \colhead{$T_\oplus$} & \colhead{$T_{bin}/T_{\oplus}$} & \colhead{$r_{\rm out}$} & \colhead{$a_{cr}$}  \\
 & \colhead{($L_\odot$)} & \colhead{(AU)} & \colhead{(yr)} & & \colhead{(AU)} & \colhead{(yr)} & & \colhead{(AU)} & \colhead{(AU)} }
\startdata
$\alpha$ Cen A & 1.519  & 0.924 & 0.834 & 95.79 & 1.233 & 1.286 & 62.14 & 2.179 & 2.72  \\
$\alpha$ Cen B & 0.500  & 0.530 & 0.391 & 204.23 & 0.707 & 0.603 & 132.55 & 1.250 & 2.60  \\
\enddata
\tablecomments{Boundary limits of the habitable zones around each star in terms of the distance (in AU), orbital period (in yr), and the period ratio relative to the binary orbital period, $T_{bin}$.  The critical outer boundary of stability (in AU) is also given based on \cite{Quarles2017b}. }
\end{deluxetable}

We begin our simulations with the first (innermost) planet having a semimajor axis, $a_1$, at the inner edge of the empirical or conservative habitable zone for $\alpha$ Cen A or $\alpha$ Cen B. Using this reference point, we explore systems expanding our iterative scheme relative to the first planet. The spacing in $\beta$ scales with $a_1$, and previous studies around single stars usually set $a_1 = 1$ AU out of convenience, but our simulations  consider a significant perturbation from the secondary star, so the choice of $a_1$ will make a difference.  The initial mean anomaly of each planet is spaced by $2 \pi j \lambda$ radians = $360 j \lambda$ degrees, where $\lambda$ is the golden ratio ($\lambda = (1 + \sqrt 5)/2 \approx 1.61803...$), $j$ denotes the planet number, with $j = 1$ corresponding to the innermost planet, and the modulo operator is used to keep the values between 0$^\circ$--$360^\circ$ (i.e., $M_1 = 222.4922^\circ$) .

\begin{figure*}
{
\plotone{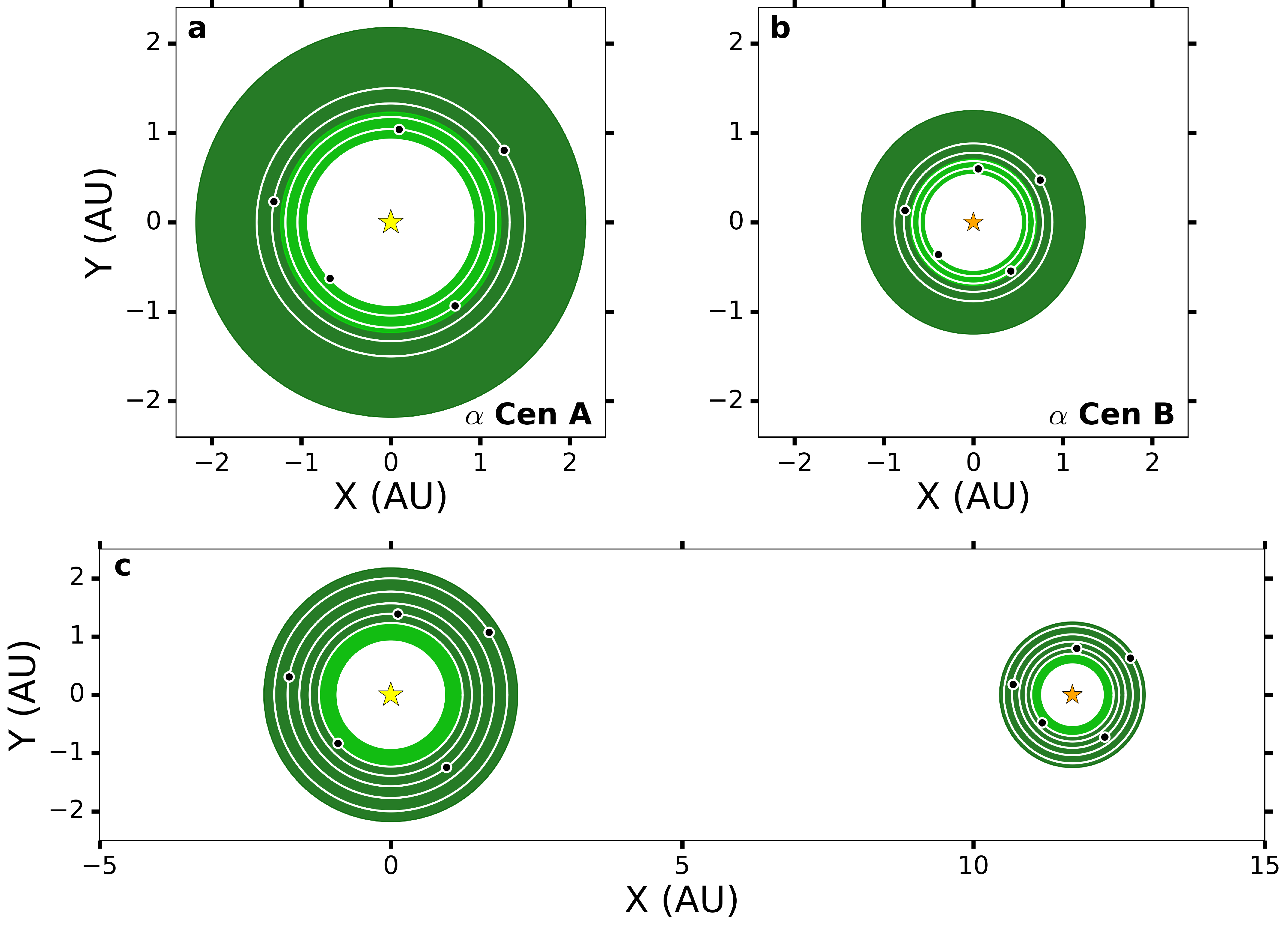}}
\caption{ Orbital schematics of the initial conditions for planets in our simulations that initially have an orbital spacing $\beta = 10$, shown from a top-down view. The empirical habitable zones, 0.32 S$_\oplus$ -- 1.78 S$_\oplus$, of each star are represented in green, with the dark green portions representing the conservative HZs, 0.32 S$_\oplus$ -- 1 S$_\oplus$. All planets orbit in the counterclockwise direction, as do the stars. Panels (a) and (b) illustrate five planet systems that begin the innermost planet at the inner edge of the host star's empirical HZ (light green).  Panel (c) illustrates similar conditions, but for the conservative habitable zone (dark green) and presents a broader view of the system including both stars' habitable zones when the stars are at periapse. \label{Schematic}}
\end{figure*}

Figure \ref{Schematic} illustrates how our orbital architectures are initially defined for $\beta = 10$, shown from a top-down view using the conservative (dark green) and empirical (light and dark green) habitable zone limits.  Figures \ref{Schematic}a and \ref{Schematic}b show the relative sizes of the empirical habitable zones resulting from differences in luminosity between the stellar components and the initial locations of five planet systems.   Using the conservative definition of the habitable zone, Figure \ref{Schematic}c illustrates that these regions are widely separated even at the time of periastron passage of the binary stars.

All of the simulations of planets orbiting $\alpha$ Cen A use the same timestep of 15.23 days, which samples the orbit of the innermost planet at least 20 times, and the timestep for simulations around $\alpha$ Cen B is chosen to be 7.15 days to account for the shorter orbital period at the inner edge of the fainter star's empirical HZ.  We iterate in the parameter $\beta$ within the 2 -- 20 range in steps of 0.05 and from 20 -- 40 in steps of 0.1.  This is adequate to probe the rise and turn-over of stability in most cases.  \cite{Obertas2016} recently probed in $\beta$ for planets around single stars at a higher resolution by drawing more than 10,000 samples from a uniform distribution and found that sampling in $\beta$ at the resolution that we are using is sufficient in the sense that increased resolution does not dramatically affect the results (i.e., by finding more stable cases in regions of overall unstable ones).   We expect that a substantial range in $\beta$ will be long-lived ($t>6$ Gyr) for systems with only a few planets, and thus we stop iterating once 10 runs have reached 10 Gyr.  

Our results will be dynamically shaped by outer boundary limits, most notably the outer edge of each star's HZ, $r_{\rm out}$, and the outer stability limit for a single planet on a circular orbit, $a_{cr}$ \citep{Wiegert1997,Holman1999,Quarles2016,Quarles2017b}.  These outer boundary limits are most easily defined with respect to a given distance in AU, and Table \ref{tab:HZ} lists the values of $a_{cr}$ in AU recently found for each star by \cite{Quarles2017b}.  {However, our study warrants describing the outer boundary limits in units of $\beta$ for each specified number of 1~M$_\oplus$ planets and each choice of the inner planet's semimajor axis.  Ignoring terms of order $m_j/M_\star$, some algebraic manipulation of Equation 9 of \cite{Obertas2016} enables us to express the value of $\beta$\footnote{The expression we derive can be further generalized to consider the spacing between any pair of planets through a substitution of indices.} for a system of $N$ planets with the inner planet at $a_1$ and outer planet at $a_N$ as:}
\begin{align}
\beta &= \left({ {\left(a_N / a_1\right)}^{1 \over N-1} - 1 \over {\left(a_N / a_1\right)}^{1 \over N-1} + 1}\right)\left({12 M_\star \over  m_j} \right)^{1/3}.
\end{align}

\noindent Table \ref{tab:outer_HZ} gives the values of the parameter $\beta$ for which the outermost planet would begin at either the outer edge of each star's HZ or the outer stability limit for a single planet on a circular orbit.

\begin{deluxetable}{cc|c|c|c}
\tablecolumns{5}
\tablewidth{0pc}
\tablecaption{Boundary Limits in Units of $\beta$ \label{tab:outer_HZ}}
\tablehead{ \colhead{Star}  & \colhead{$a_1$} & \colhead{$n_{pl}$} & \colhead{$r_{\rm out}$ $(\beta)$} & \colhead{$a_{cr}$ $(\beta)$} }
\startdata
{\multirow{8}{*}{\rotatebox[origin=c]{90}{$\alpha$ Cen A}}} & {\multirow{4}{*}{$r_{\rm emp}$}}& 2 & 66.90 & 81.53 \\
&& 3 & 34.95 & 43.59 \\
&& 4 & 23.50 & 29.44 \\
&& 5 & 17.68 & 22.20 \\
\cline{2-5}
& {\multirow{4}{*}{$r_\oplus$}} & 2 & 45.86 & 62.23 \\
&& 3 & 23.39 & 32.29\\
&& 4 & 15.65 & 21.68 \\
&& 5 & 11.75 & 16.31\\
\hline
{\multirow{8}{*}{\rotatebox[origin=c]{90}{$\alpha$ Cen B}}} & {\multirow{4}{*}{$r_{\rm emp}$}}& 2 & 63.58 & 103.95 \\
&& 3 & 33.21 & 59.39 \\
&& 5 & 16.79 & 30.84 \\
&& 6 & 13.45 & 24.79 \\
\cline{2-5}
& {\multirow{4}{*}{$r_\oplus$}} & 2 & 43.61 & 89.98 \\
&& 3 & 22.24 & 49.43\\
&& 5 & 11.18 & 25.36 \\
&& 6 & 8.95 & 20.36
\enddata
\tablecomments{ Outer edge, $r_{\rm out}$, of the HZs and stability limit, $a_{cr}$, for initially circular orbits, given in units of the planetary spacing parameter $\beta$.  This boundary depends on the host star, the initial semimajor axis of the innermost planet, $a_1$, and the number of planets in the system, $n_{pl}$.} 
\end{deluxetable}

\subsection{Free and Forced Eccentricity}

\begin{figure*}
\epsscale{1.}
\plotone{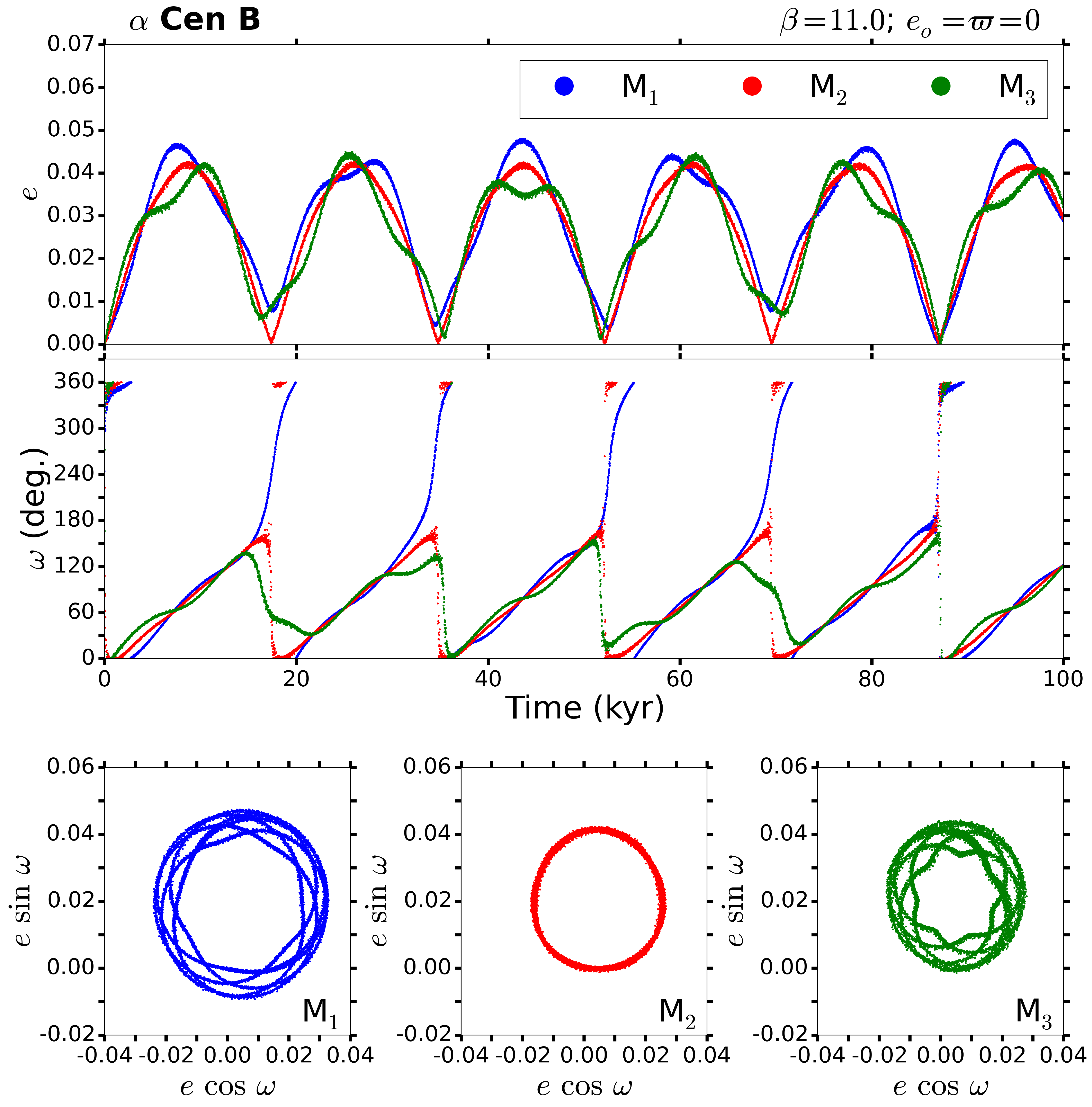}
\caption{Evolution of an initially \emph{circular} three planet system orbiting $\alpha$ Cen B over the first 100,000 yr, where the innermost planet begins at $r_{emp}$.  The top panel shows the variation of the eccentricities of each of the planets, and the middle panel illustrates the changes in the argument of periastron, $\omega$.  These orbital elements couple together in the bottom panel, which presents the eccentricity vector taken by each planet during this time interval.  \label{ap_circ}}
\end{figure*}

\begin{figure*}
\epsscale{1.}
\plotone{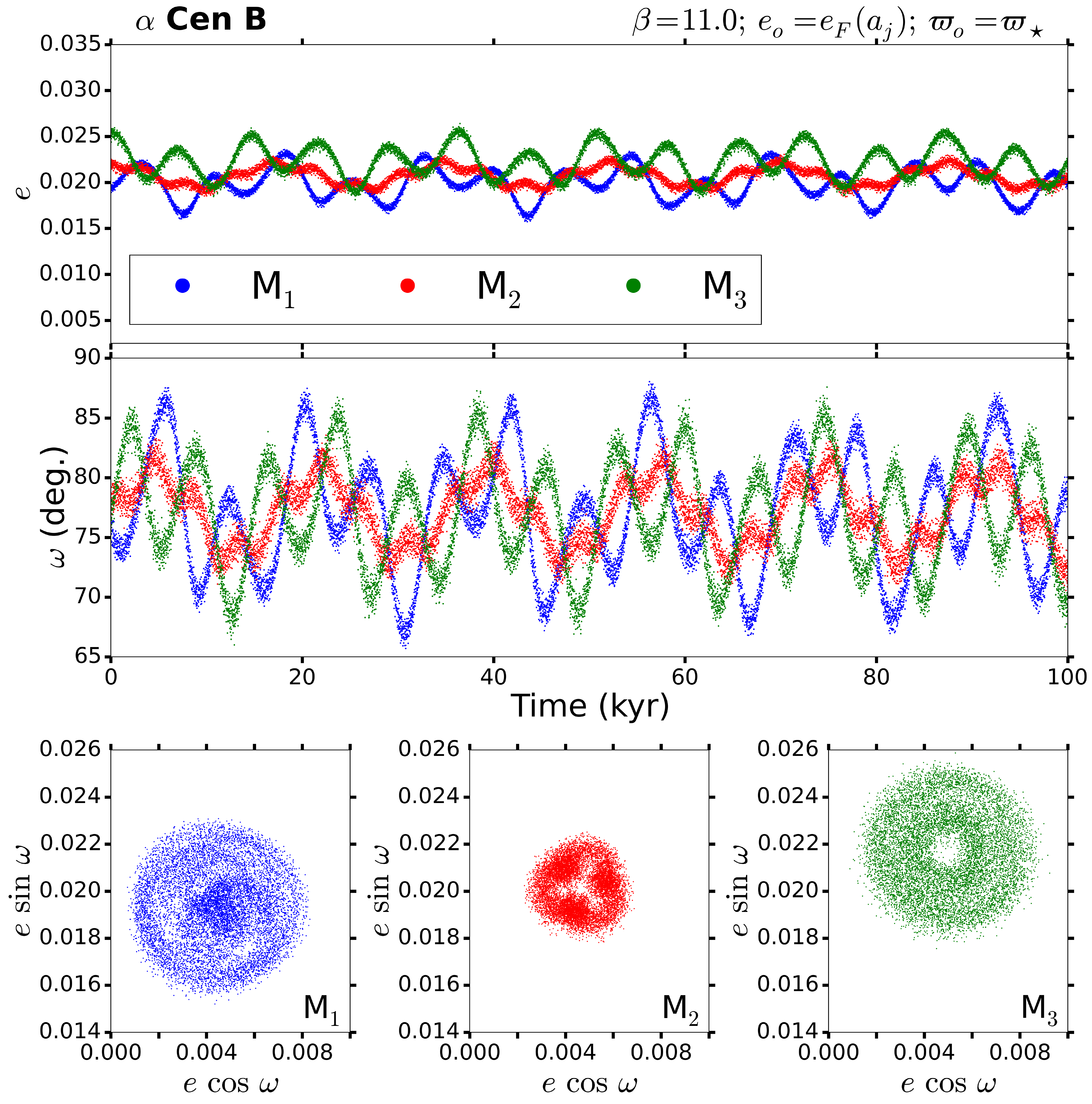}
\caption{Early evolution of a three planet system orbiting $\alpha$ Cen B, where the innermost planet begins at $r_{emp}$ with initial planetary \emph{eccentricities} approximately equal to their forced eccentricity, thereby minimizing free eccentricities.  The top panel shows the variation of eccentricity between the planets, and the middle panel illustrates the changes in the argument of periastron, $\omega$.  These orbital elements couple together in the bottom panel, which presents the eccentricity vectors that each planet takes during this interval of time.  \label{ap_ecc}}
\end{figure*}

A planet beginning on a circular orbit in the HZs of $\alpha$ Cen A or B rapidly reaches an eccentricity of $\sim$~0.05 through eccentricity pumping from the binary companion, but planets assigned their forced eccentricities show smaller oscillations in the total eccentricity.  The maximum eccentricity obtained in the initially circular runs is close to $2 e_F$, where $e_F(a)$ is the eccentricity that is forced by the companion star on a planet orbiting with semimajor axis $a$. {Apsidal precession of the pericenter also occurs in the (initially) circular runs, allowing planetary orbits to rotate from a non-crossing to a crossing orientation.  These large eccentricities allow planets to strongly perturb one another, in many cases leading to collisions.  Figure \ref{ap_circ} illustrates how a three planet system around $\alpha$ Cen B with $\beta = 11.0$ can evolve in eccentricity and periastron angle over a timescale of 100,000 years.  The oscillation timescale appears to be $\sim$18,000 years, and  the inner planet's periapse angle circulates with this period.  Figure \ref{ap_ecc} illustrates a similar system to that represented in Figure \ref{ap_circ}, but with assigned initial eccentricity equal to $e_F(a)$  for each of the planets.  In this case, the variations in eccentricity and periapse angles are much smaller.}

Processes during planet formation tend to drive planetary systems towards a minimum in free eccentricity. This favors circular orbits in single star systems, but can lead to eccentric orbits in both the circumbinary environment \citep{Bromley2015} and the circumstellar environment when the host star has a more distant stellar companion.  These two facts motivate us to explore the difference in the smallest stable value of $\beta$ both for planets with initially circular orbits and for planets with initial eccentricities chosen so that they begin with minimized free eccentricity.
In order to supply a good estimate of $e_F$ to use as the initial eccentricity for this second set of simulations, we build upon the numerical method described in \cite{Andrade2016} and the results of \cite{Quarles2017b} for prograde bodies.  We use the results from \cite{Quarles2017b} that includes a piecewise-quadratic formula, which adequately provides a forced eccentricity $e_F$ as a function of the starting semimajor axis for the $j^{th}$ planet $a_j$ in our standard ``eccentric'' simulations.  This formula is applied to orbits around each star in the binary system using the form of $e_F = C_1a_j^2 + C_2a_j + C_3$, whose coefficients are given in Table \ref{tab:coeff} for each star.  Note that the difference between coefficients for the two stellar components are small because the stars have similar masses.

\begin{deluxetable}{ccccc}
\tablecolumns{5}
\tablewidth{0pc}
\tablecaption{Coefficients for Forced Eccentricity \label{tab:coeff}}
\tablehead{ & \colhead{Star} &\colhead{$C_1$} & \colhead{$C_2$} & \colhead{$C_3$}}
\startdata
$a<2$ AU & $\alpha$ Cen A & --0.007  & 0.044 &  --0.002 \\
& $\alpha$ Cen B & --0.009 & 0.047 & --0.003 \\
$a \geq 2$ AU & $\alpha$ Cen A & --0.027  & 0.123 &  --0.080 \\
& $\alpha$ Cen B & --0.030 & 0.130 & --0.085 \\
\enddata
\tablecomments{Coefficients for determining the forced eccentricity, $e_F$, for planets on prograde orbits as a quadratic function in starting semimajor axis around each stellar component \citep{Quarles2017b}.}
\end{deluxetable}

\subsection{Two, Three, and Five Planet Systems}
We perform a series of simulations considering two, three, and five planet systems around either star in the $\alpha$ Cen system.  The inner planet in each of these runs begins at a semimajor axis that corresponds to the inner edge of either the empirical HZ ($r_{\rm emp}$) or the conservative HZ ($r_\oplus$) of its host star, and we only iterate in $\beta$ until the first ten runs that survive 10 Gyr are reached.  The length of our simulations extend beyond our criterion for stability (6 Gyr) to account for extremely late instabilities (i.e., those that can occur in the future), as the Solar System can be considered unstable as per our definition \citep{Batygin2008,Batygin2015,Laskar2009} and there remains some uncertainty in the age of the $\alpha$ Cen AB system \citep{Mamajek2008}.  In addition, our choice in the initial eccentricity of the planets may affect our results, so we explore a similar set of simulations where each of the planets begins with an eccentricity close to the forced eccentricity at its location in the binary star system.

In most cases, our three and five planet runs broadly explore up to $\beta = 40$ to show the full range of outcomes.  Lifetimes generally increase with $\beta$ and then decrease once the starting semimajor axis of the outermost planet approaches the stability limit for single planet systems in the $\alpha$ Cen binary star system \citep{Wiegert1997,Holman1999,Quarles2016}, where the specific values of $\beta$ are given in Table \ref{tab:outer_HZ}.

\begin{figure*}
\includegraphics[width=\textwidth]{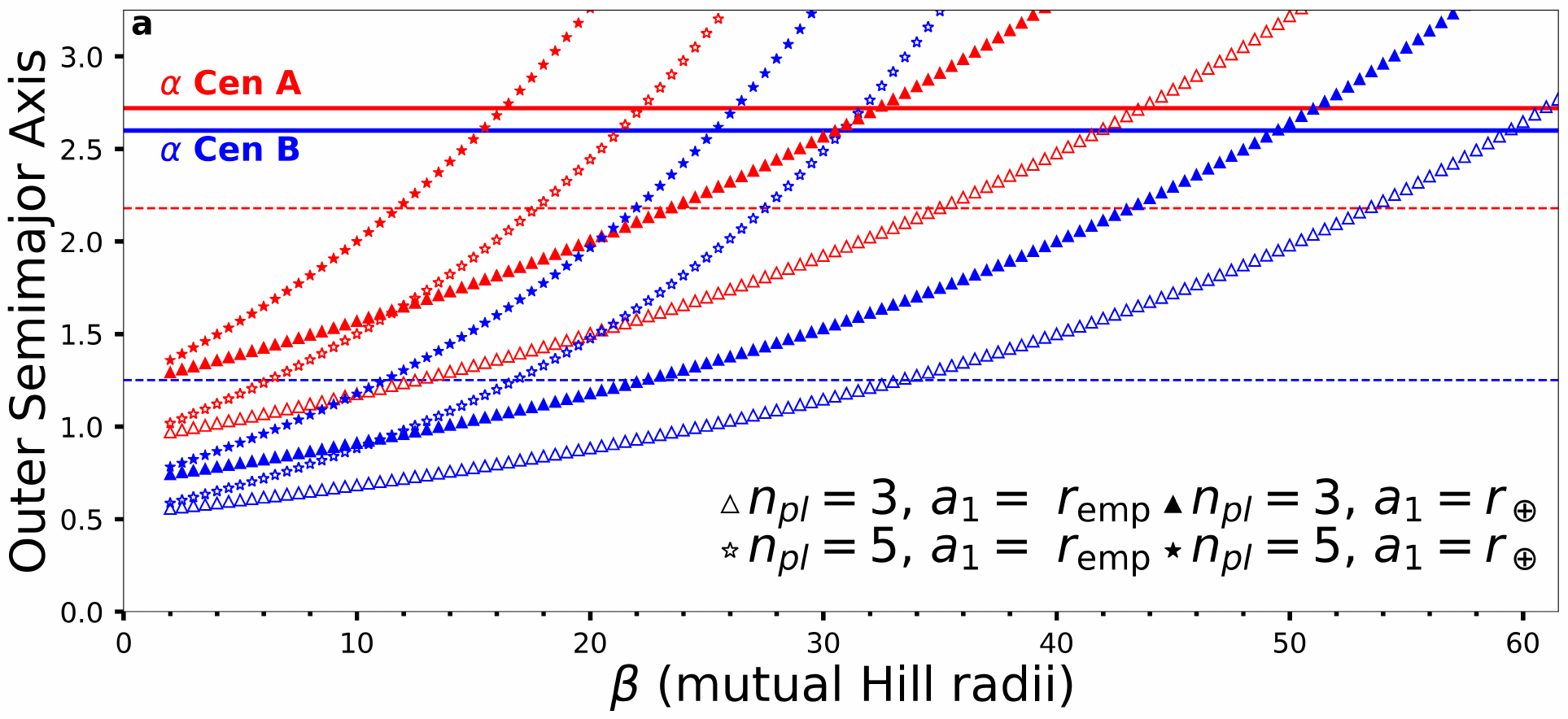}
\includegraphics[width=\textwidth]{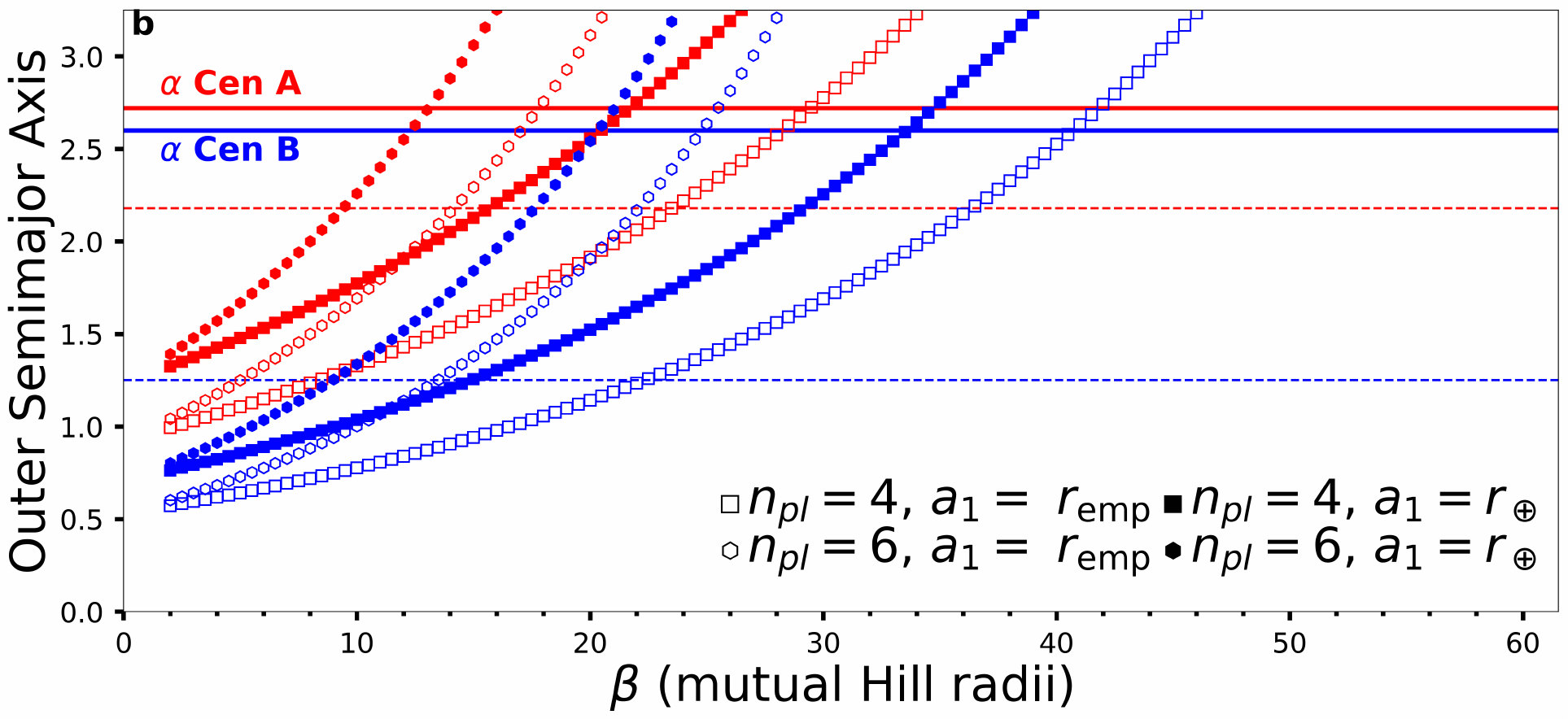}
\caption{The starting semimajor axis of the outermost planet for three (triangles), four (squares), and five (stars) planets orbiting $\alpha$ Cen A or three (triangles), five (stars), and six planets (hexagons) orbiting $\alpha$ Cen B relative to the spacing parameter $\beta$, where the starting semimajor axis begins at the inner edge of the empirical (open symbols) or conservative (filled symbols) HZ.  The outer edge of the HZ for each stellar host is represented by the dashed (red and blue) horizontal lines.  The solid (red and blue) horizontal lines denote the stability limits for single planets on prograde orbits about $\alpha$ Cen A and $\alpha$ Cen B, respectively, using those established in \cite{Quarles2017b}. \label{outer_SA}}
\end{figure*}

We plot the initial semimajor axis of the outermost planet, $a_{3}$ or $a_{5}$,  as a function of $\beta$ in Figure \ref{outer_SA}a, to show how the single-planet stability limit affects the exploration in $\beta$.  The points in Figure \ref{outer_SA}a are delineated by color (red or blue) representing the host star, by symbol (dot or triangle) denoting the number of planets in the system, and by the innermost semimajor axis beginning at the inner edge of the empirical (open) or conservative (filled) HZ of each star.  The stability limit (red solid line) for five planets (red open triangles) in the conservative HZ orbiting $\alpha$ Cen A is crossed at $\beta \approx 21$, whereas the three planet (red open dots) cases crosses the stability boundary at $\beta \approx 41$ (See Table \ref{tab:outer_HZ} for more precise values.).  The outer edge of the HZ (dashed) is also indicated, showing that a range of values in $\beta$ may be allowed through stability without all the planets residing within the respective HZ of the host star. Some of our simulations include planets exterior to the HZ, making our numerical results applicable to any choice for the outer boundary of the HZ.  

\subsection{Four and Six Planet Systems}
One of the primary goals of our simulations is to determine how many planets can reside within the HZs of $\alpha$ Cen A and B.  Therefore, in addition to our general study of two, three, and five planet systems, we also perform a limited study of four planet systems with a forced eccentricity around $\alpha$ Cen A and six planet systems with a forced eccentricity in $\alpha$ Cen B.  {The motivation for including these specific systems is largely informed by our results in Section \ref{sec:ecc_orbits}, as discussed in the first paragraph of Section \ref{sec:46plan}.}  We have limited our investigation to $\beta=25$, as the peak in lifetime typically occurs at a smaller value of $\beta$ than this.  We show in Figure \ref{outer_SA}b the location of the outer planet as a function of $\beta$ and also delineate the stability limit and the outer edge of each star's respective HZ, $r_{\rm out}$. (See Table \ref{tab:outer_HZ} for specific values.)

\section{Results}
The results of our simulations are plotted in Figures \ref{235_A} -- \ref{235_B_ecc} and \ref{46_ecc}.  We use the common convention of computing the lifetime of the initial configuration (which we plot on a logarithmic scale) as a function of the planetary spacing parameter $\beta$.  We delineate our results using a color-code, where black dots signify those runs with $a_1$ set to the inner edge of the empirical habitable zone, $r_{\rm emp}$, and red dots represent runs with $a_1$ set to an orbit where the planet intercepts the same flux of radiation as does the Earth, $r_\oplus$, i.e., the inner edge of our conservative HZ.  We refer to a configuration as ``stable'' when its lifetime extends beyond 6 Gyr. To guide the eye, we have indicated the stable runs by highlighting them with upwards vertical ticks, retaining the color-code to signify the location of the innermost planet, and used open symbols to designate those runs that survived the entire 10 Gyr simulated.  

\subsection{Two, Three, and Five Planet Systems} \label{sec:235plan}
These runs explore the lifetime of nested coplanar orbits of two, three, or five Earth-mass planets orbiting either $\alpha$ Cen A or $\alpha$ Cen B.  The stability of two planet systems has been well-studied around single Sun-like stars \citep{Gladman1993,Chambers1996}, but has not been extensively explored in the case of binary star systems.  The presence of a second star changes the character of the problem because the gravitational potential of the many-body system can be substantially different from one in which only three bodies are present, allowing for more degrees of freedom in the system and thereby preventing systems from becoming Hill stable. 

The lifetime verses spacing plots share many characteristics with analogous plots for systems of three or more planets around single stars  \citep{Smith2009,Obertas2016}.  For closely spaced systems, the logarithm of system lifetime increases roughly in proportion to planetary spacing, with shorter lifetimes near strong mean motion resonances and scatter caused by chaos.  However, systems tend to be shorter-lived than comparable systems orbiting single stars.  And for configurations in which the outermost planet is strongly perturbed by the binary companion, system lifetime can flatten out with increased spacing and then turn over, with lifetime decreasing as planetary spacing is increased further.

\subsubsection{Circular Orbits}
\begin{figure*}
\epsscale{1.}
\plotone{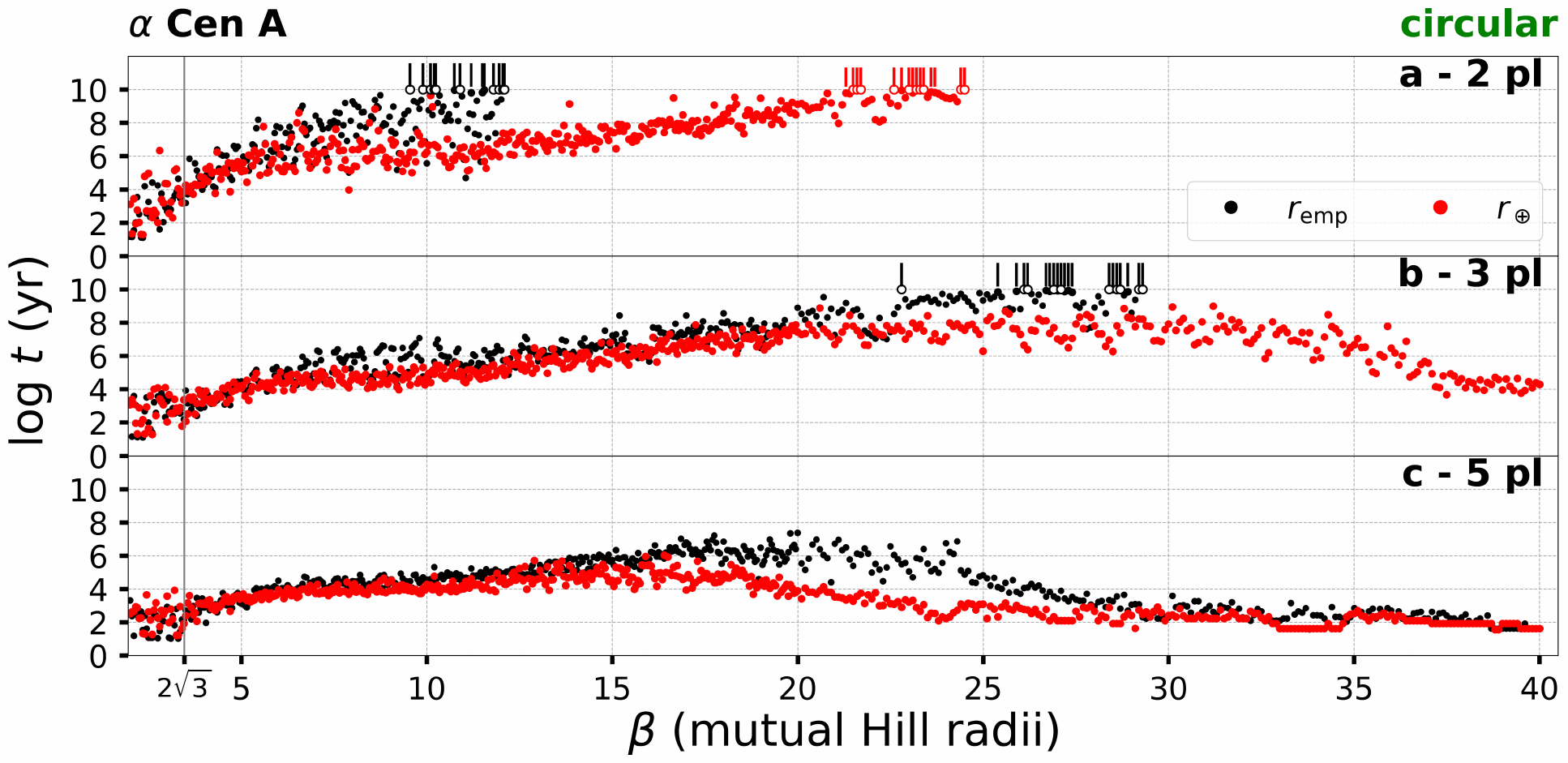}
\caption{ Lifetimes of initially circular two (a), three (b), and five (c) planet systems that orbit $\alpha$ Cen A are shown as a function of the interplanetary spacing parameter $\beta$.  Black (red) points represent those systems that begin at the inner edge of the empirical (conservative) habitable zone.  Open points represent systems that survived for the entire 10 Gyr time interval simulated.  Simulations where the survival time $t$ exceeds 6 Gyr are highlighted with vertical ticks in the space beyond 10 Gyr. A vertical (gray) line marks the two planet single star stability limit, $\beta = 2\sqrt{3}$. \label{235_A}}
\end{figure*}

\begin{figure*}
\epsscale{1.}
\plotone{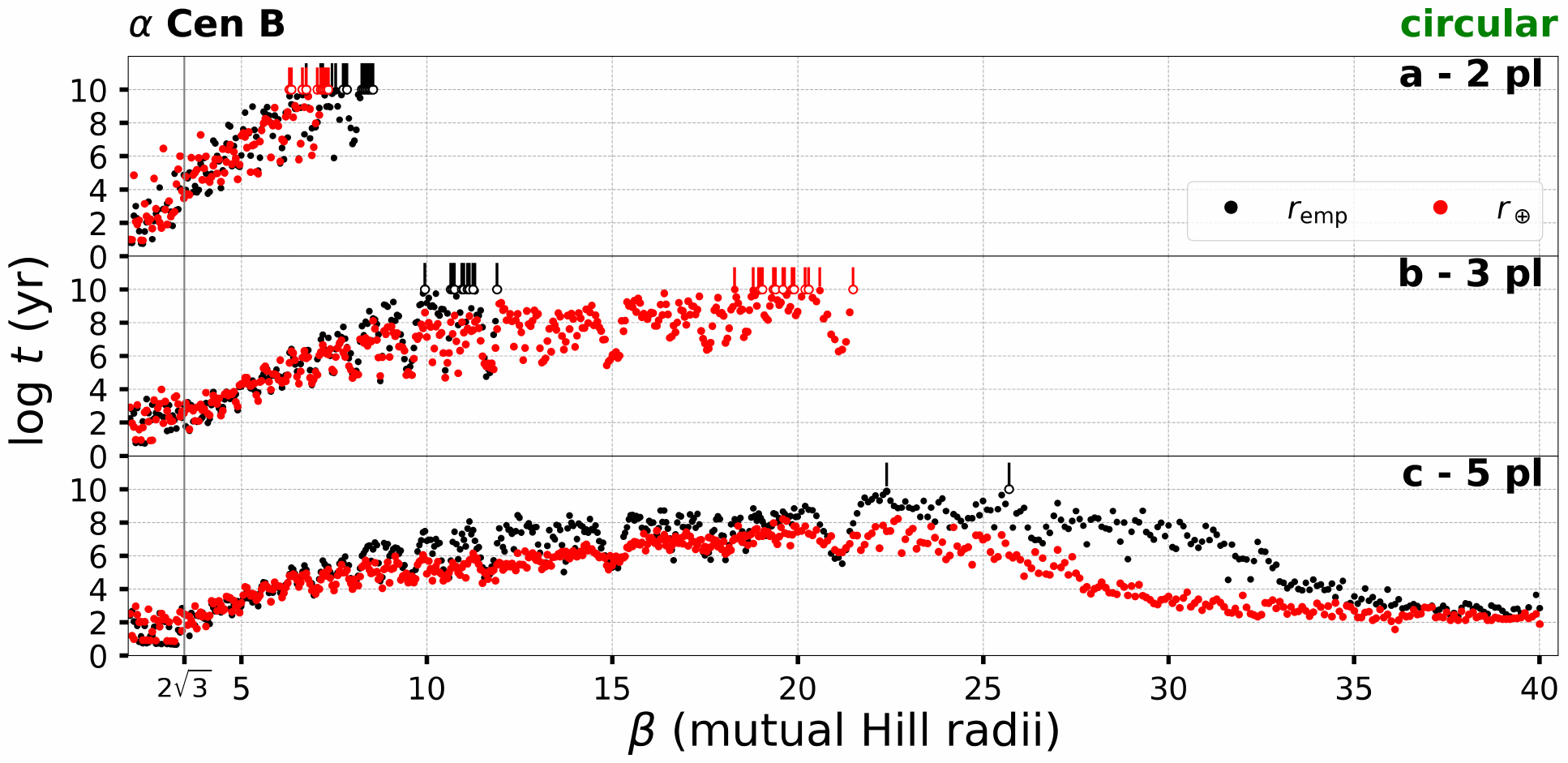}
\caption{ Lifetime of two (a), three (b), and five (c) initially circular planet systems that orbit $\alpha$ Cen B as a function of the interplanetary spacing parameter $\beta$.  See caption to Figure \ref{235_A} for an explanation of the symbols used on this plot. \label{235_B}}
\end{figure*}

We plot the lifetimes of initially circular, coplanar systems around $\alpha$ Cen A and $\alpha$ Cen B in Figures \ref{235_A} and \ref{235_B}, respectively.  Overall, Figures \ref{235_A} and \ref{235_B} show several common features.  The transition to stability occurs at substantially higher values of $\beta$ than in the case of single stars.  The additional small perturbations from the secondary star cause the location of the transition to stability to depend on the starting semimajor axis of the innermost planet ($r_\oplus$ or $r_{\rm emp}$). As is the case for multi-planet systems orbiting single stars \citep{Smith2009,Obertas2016}, interplanetary perturbations due to mean motion resonances between the planets can reduce the lifetimes of systems by up to several orders of magnitude, resulting in stable islands in $\beta$ between these resonances.  For $\beta > 6$, simulations with the innermost planet beginning near the inner edge of the empirical HZ ($r_{\rm emp}$) typically survive for longer times than those runs with the innermost planet initially near the inner edge of the conservative HZ ($r_\oplus$).  Also, systems of planets orbiting $\alpha$ Cen A are typically shorter-lived at a given value of $\beta > 5$ than systems with the same number of planets that intercept the same radiation flux around $\alpha$ Cen B, because the difference between the luminosities of the stars is much larger than the difference in their masses.

Figure \ref{235_A}a illustrates that two planet systems around $\alpha$ Cen A with initially circular orbits are much more stable if $a_1 = r_{\rm emp}$ than if  $a_1 = r_{\oplus}$.  Stable orbits are possible for relatively tightly spaced ($\beta = 9.55$) pairs of planets near the inner boundary of the emperical HZ.  In contrast, much more widely  spaced orbits ($\beta > 21$) are necessary for stability by our definition (survival for $> 6$~ Gyr) if the innermost planet begins at 1 Earth flux, although, several planet pairs that are much more closely spaced are long-lived, with the system at $\beta = 10$ surviving for $> 4.5$~ Gyr. 

The three planet runs in Figure \ref{235_A}b show stable configurations are achieved only for $r_{\rm emp}$ beginning at $\beta \approx 23$, and all of the three planet $r_\oplus$ runs undergo collisions between the planets within 1 Gyr.  Finally, the five planet systems in Figure \ref{235_A}c become unstable on significantly shorter timescales, where the longest lifetimes are less than 100 Myr.  This is largely due to outermost planet beginning close to or beyond the outer stability limit even for the smallest values of $\beta$ for which three planet systems are stable with the innermost planet located at $r_{\rm emp}$; see Figure \ref{outer_SA}a and Table \ref{tab:outer_HZ}.  

The initially circular, coplanar systems around $\alpha$ Cen B whose lifetimes are displayed in Figure \ref{235_B} show a behavior intermediate between the systems around $\alpha$ Cen A displayed in Figure \ref{235_A} and planetary systems orbiting the Sun.  The minimum separation for two planets to be stable is $\beta \sim 6.5$ (Figure \ref{235_B}).  Three planets become stable at larger values ($\beta \sim 10$) for the $r_{\rm emp}$ cases as shown in Figure \ref{235_B}b, somewhat larger than the $\beta \sim 6.4$ minimum spacing of comparable systems around single stars found by \cite{Smith2009}.  For the $r_\oplus$ runs, stable systems occur when $\beta > 18$.  Only two of the five planet systems are long-lived in Figure \ref{235_B}c and both have the starting semimajor axis of the innermost planet located at $r_{\rm emp}$.  We note that even though stable configurations exist, only four of the five planets reside interior to the outer boundary of our defined habitable zone.  

\subsubsection{Orbits with Minimal Free Eccentricity} \label{sec:ecc_orbits}
A single planet initially on a circular orbit about an isolated star will remain on the same orbit indefinitely (neglecting stellar evolution and gravitational radiation), but an eccentric binary companion in a system like $\alpha$ Cen will excite a forced eccentricity.  Therefore, we simulate systems of two, three or five planets around each of the stellar components with a forced eccentricity, $e_F$, that depends on the starting semimajor axis of each planet in a simulation.  All of the planets begin with a common argument of periastron that is aligned with the binary orbit, so that $\Delta \varpi = 0^\circ$.    

\begin{figure*}
\epsscale{1.}
\plotone{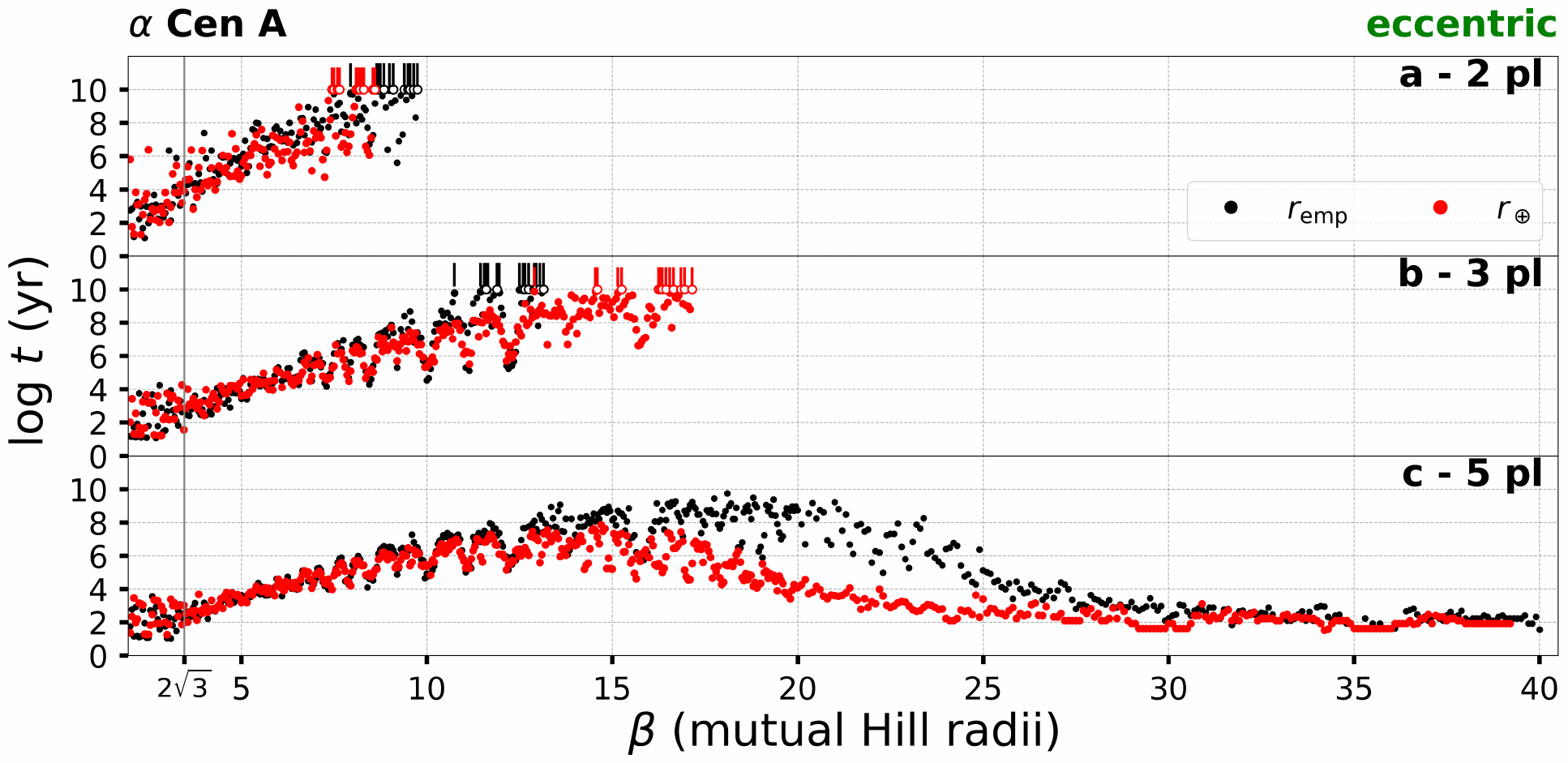}
\caption{Lifetime of two (a), three (b), and five (c) initially eccentric planet systems that orbit $\alpha$ Cen A, with eccentricity chosen to minimize free eccentricity, as a function of the interplanetary spacing parameter $\beta$.  
See caption to Figure \ref{235_A} for an explanation of the symbols used on this plot. \label{235_A_ecc}}
\end{figure*}

\begin{figure*}
\epsscale{0.55}
\plotone{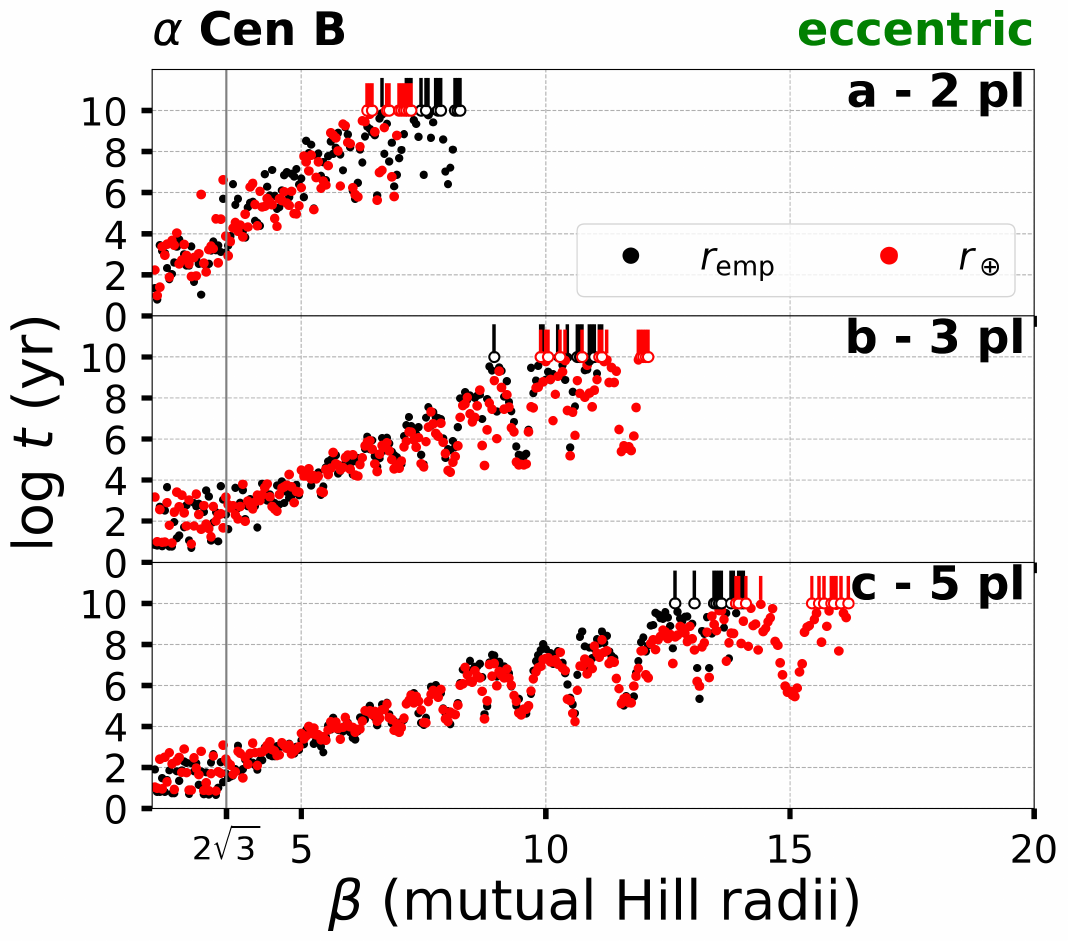}
\caption{ Lifetime of two (a), three (b), and five (c) initially eccentric planet systems that orbit $\alpha$ Cen B as a function of the interplanetary spacing parameter $\beta$.  See caption to Figure \ref{235_A} for an explanation of the symbols used on this plot. \label{235_B_ecc}}
\end{figure*}

Planetary systems around $\alpha$ Cen A with minimized initial free eccentricity are stable for much tighter configurations than those begun on circular orbits, as can be seen by comparing Figures \ref{235_A_ecc} and \ref{235_B_ecc} to Figures \ref{235_A} and \ref{235_B}. The most tightly packed long-lived ($>6$ Gyr) three planet runs occur at $\beta \sim 11$ for $e_o = e_F$, in contrast to $\beta \sim 23$ for $e_o = 0$.  In Figure \ref{235_A_ecc}c, a region exists between 16 and 21 in $\beta$ that allows for five planet configurations with $a_1 = r_{\rm emp}$ to survive past 1 Gyr, although they do not last up to 6 Gyr to meet our stability criterion.  As for systems of planets with $e_o = 0$, some regions in $\beta$ are dominated by chaos or mean motion resonances.  However,  lifetimes of systems with a forced eccentricity are less sensitive to the starting semimajor axis of the innermost planet $a_1$, as shown by a smaller deviation between runs at a constant $\beta$. 

For two planets initially orbiting $\alpha$ Cen B, Figure \ref{235_B_ecc}a shows the minimum long-lived $\beta$ is nearly the same value ($\sim 6.7$) as compared to Figure \ref{235_B}a.  But the long-lived three planet runs (vertical ticks), appear at a smaller value in $\beta$ for minimized free eccentricity (Figure \ref{235_B_ecc}b) than those for initially circular orbits (Figure \ref{235_B}b), with the difference being much larger for systems beginning at $r_\oplus$ than for those with inner planet at $r_{\rm emp}$.  This is also the case in the five planet runs, but to a much stronger degree, with a difference of $\sim 10$ in the smallest stable $\beta$ for $a_1 = r_{\rm emp}$, and five planets can even survive within the conservative HZ.     

\begin{figure}
\epsscale{0.55}
\plotone{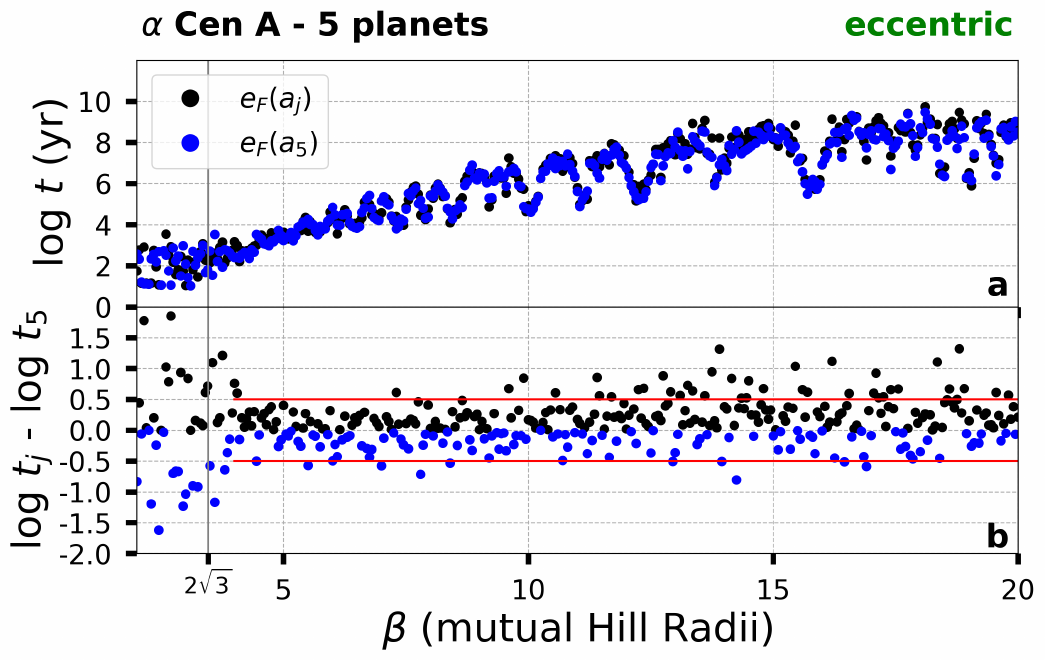}
\caption{ Comparison of lifetimes of systems wherein the initial eccentricities of all the planets are set to the forced eccentricity at the semimajor axis of the outermost planet, $e_F(a_5)$, (blue)  to those with initial eccentricity of each planets is set to the forced eccentricity at its own  semimajor axis, $e_F(a_j)$, (black; plotted in panel (a) using the same data used for the black points in Figure \ref{235_A_ecc}c) for five planets initially orbiting $\alpha$ Cen A.  Panel (a) illustrates the similarities between the runs, and panel (b) shows at an expanded vertical scale the pointwise differences between the runs, with black points representing values of $\beta$ where the system with $e_F(a_j)$ is longer lived and blue points representing values of $\beta$ where the system with $e_F(a_5)$ survives longer.  Horizontal red lines are plotted to guide the eye and differentiate between regions of random scatter and those that could be more significant.  A vertical (gray) line marks the two planet single star stability limit,  $\beta = 2\sqrt{3}$.  \label{ecsame}}
\end{figure}

Our algorithm for  choosing eccentricities of each planet independently leads to different eccentricities for each planet and thus to non-zero initial \emph{relative} eccentricities of the planets.  One might think that starting all of the planets with the same eccentricity, set by our estimate of the forced eccentricity of the outermost (and most perturbed) planet, might lead to longer system lifetimes.  We thus re-ran the simulations of five planets systems with the inner planet orbiting $\alpha$ Cen A at $r_{\rm emp}$ with all planetary eccentricities initially equal to the forced eccentricity of the outermost planet.  As shown in Figure \ref{ecsame}, on average, this prescription typically resulted in slightly \emph{shorter} system lifetimes.

\subsection{Four and Six Planet Systems} \label{sec:46plan}
Our three planet simulations around $\alpha$ Cen A that minimize free eccentricity show stable values of $\beta$ with the semimajor axis of the outermost planet occurring well inside the outer limit of the HZ and the stability limit for single planets (Figure \ref{235_A_ecc}b), suggesting that four planet systems might be stable within the HZ of $\alpha$ Cen A.  Similarly, our five planet simulations (Figure \ref{235_B_ecc}c) suggest that six planets might be stable within the HZ of $\alpha$ Cen B.  Thus, we also explore the stability of four planet systems around $\alpha$ Cen A and six planet systems around $\alpha$ Cen B.  We note that the outer planet is located at the single planet stability limit for a lower value of $\beta$ when the number of planets is larger (see Figure \ref{outer_SA}b and Table \ref{tab:outer_HZ}).  

\begin{figure*}
\epsscale{1.}
\plotone{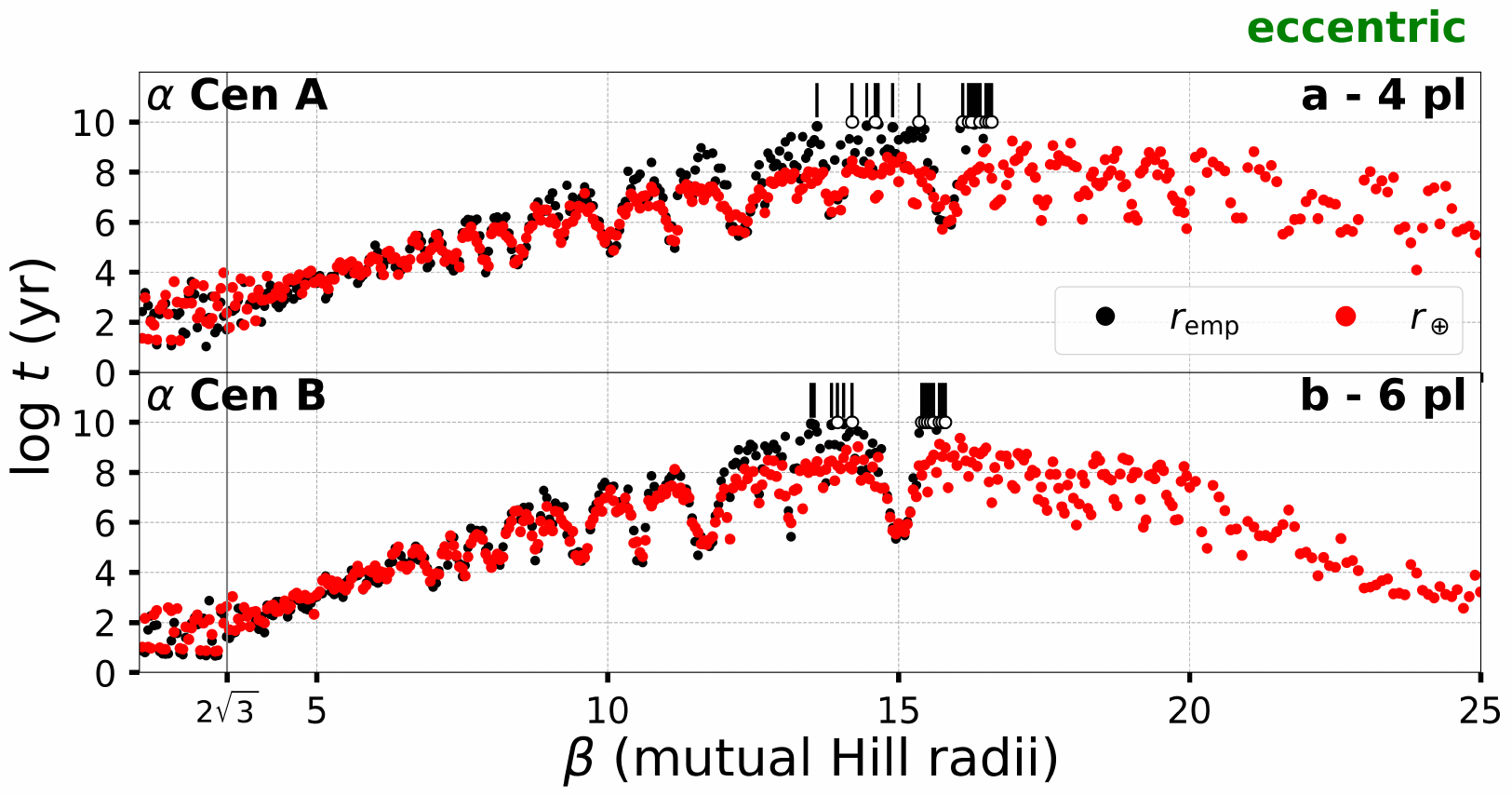}
\caption{ Lifetime of four (a) and six (b) initially eccentric planetary systems that orbit $\alpha$ Cen A or $\alpha$ Cen B, respectively, as a function of the interplanetary spacing parameter $\beta$.  See caption to Figure \ref{235_A} for an explanation of the symbols used on this plot. \label{46_ecc}}
\end{figure*}

Figure \ref{46_ecc}a shows that four planet configurations beginning with $r_{\rm emp}$ can be stable around $\alpha$ Cen A starting near $\beta = 13.6$, well inside the outer edge of the HZ.  However, the lifetimes of systems with $a_1 = r_\oplus$ are typically lower than those with $a_1 = r_{emp}$ for $\beta \gtrsim 9$. Few simulations systems with $a_1 = r_\oplus$ survive beyond 1 Gyr  and none last long enough to be deemed stable by our definition.  None of the five planet runs in Figure \ref{235_A_ecc}c are stable for either choice of the innermost planet's semimajor axis.

The inner edge of the HZ of $\alpha$ Cen B is much closer to the host star, possibly allowing for more planets to be stable within it.  However, the width of the HZ is smaller than around $\alpha$ Cen A, albeit only slightly smaller when measured in units of $\beta$ for Earth-mass planets, and the outer edge of $\alpha$ Cen B's HZ is well interior to the largest stable prograde orbit of a single planet around $\alpha$ Cen B found by \cite{Quarles2016}.  More planets could stably orbit the star, but not all the planets would necessarily fit within the boundaries of our prescribed HZs.  In a similar fashion to Figure \ref{46_ecc}a, Figure \ref{46_ecc}b indicates that six planet systems in the $r_{\rm emp}$ runs are stable starting at $\beta = 13.50$, which places the outermost planet just beyond $r_{\rm out} (\beta) = 13.45$ (see Table \ref{tab:outer_HZ}), i.e., the outer planet orbits just exterior to the outer edge of the HZ.  None of the six planet systems that began with $a_1 = r_\oplus$ survive up to our 6 Gyr threshold for stability, and those that lasted the longest had multiple planets orbiting exterior to the HZ.

\subsection{Mean Motion Resonances}
\begin{figure*}
\epsscale{1.}
\plotone{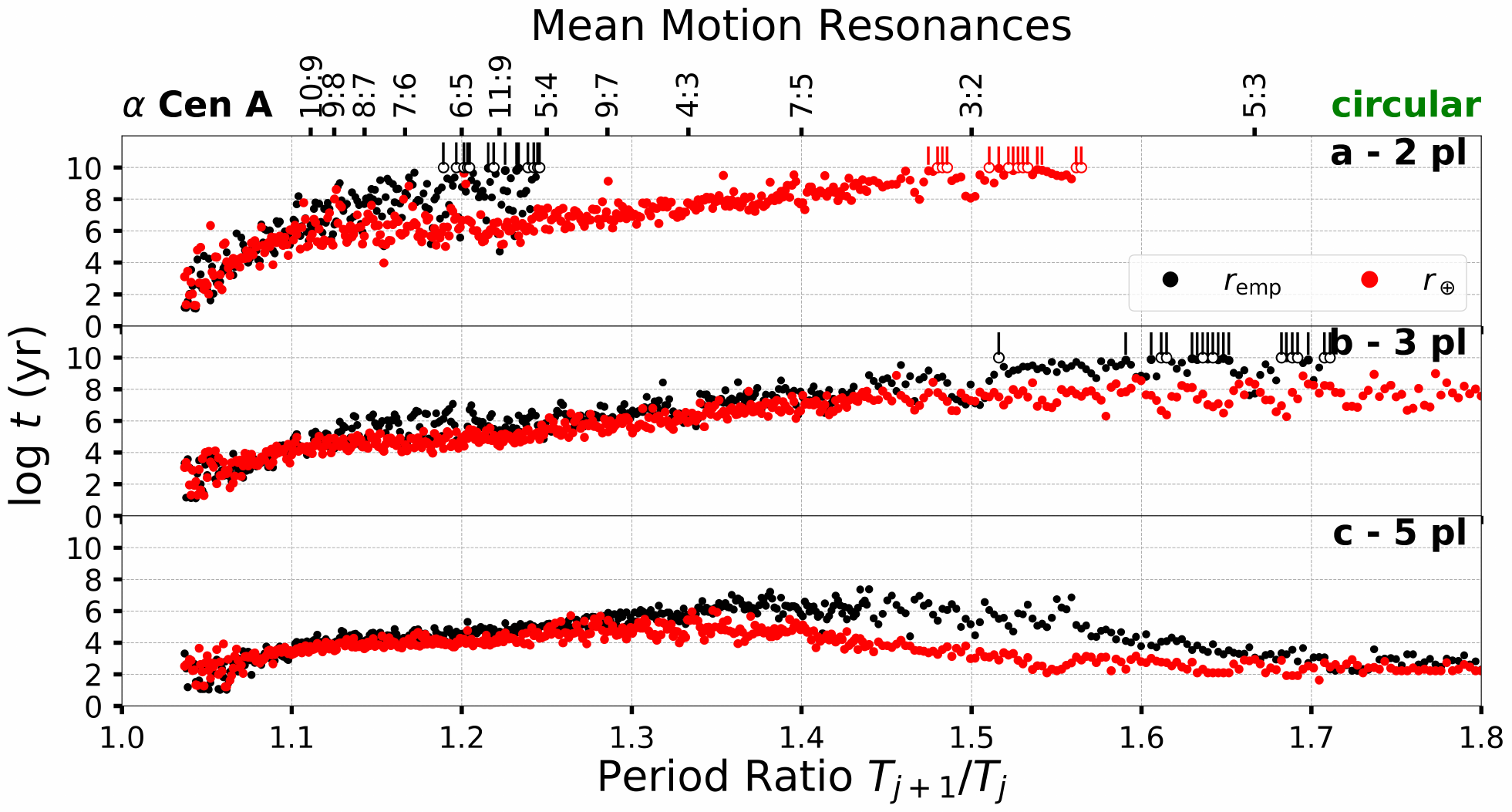}
\caption{Data shown in Figure \ref{235_A}, but the horizontal axis has been re-scaled to measure the planetary spacing in terms of the ratio of the initial orbital periods of adjacent pairs of planets.  The locations of first- and second-order mean motion resonances of adjacent planets are marked along the top axis. See caption to Figure \ref{235_A} for an explanation of the symbols used.\label{235_res}}
\end{figure*}

\begin{figure*}
\epsscale{1.}
\plotone{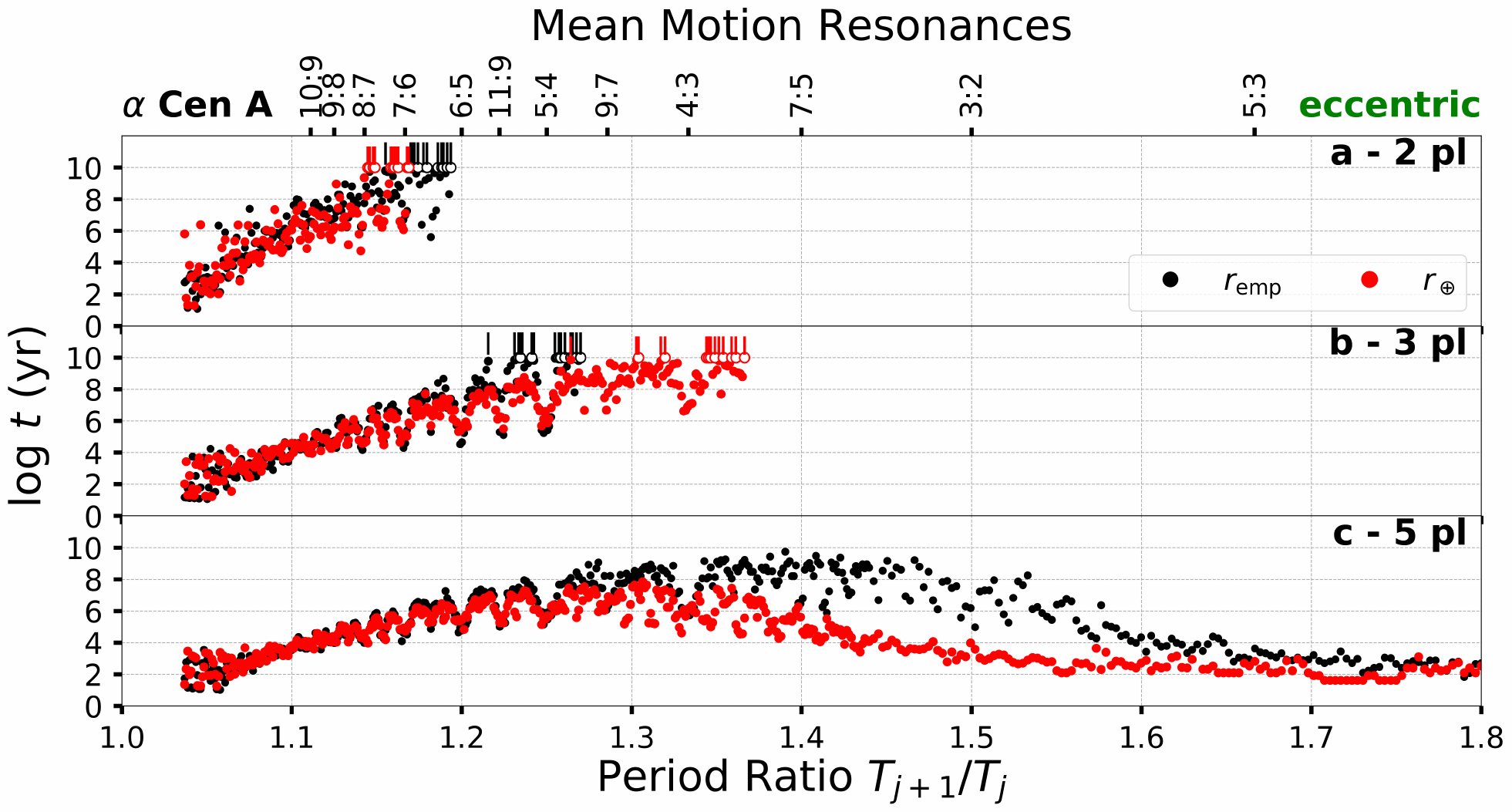}
\caption{Data shown in Figure \ref{235_A_ecc}, but the horizontal axis has been re-scaled to measure the planetary spacing in terms of the ratio of the initial orbital periods of adjacent pairs of planets.  The locations of first- and second-order mean motion resonances of adjacent planets are marked along the top axis. See caption to Figure \ref{235_A} for an explanation of the symbols used.\label{235_res_ecc}}
\end{figure*}

\begin{figure*}
\epsscale{1.}
\plotone{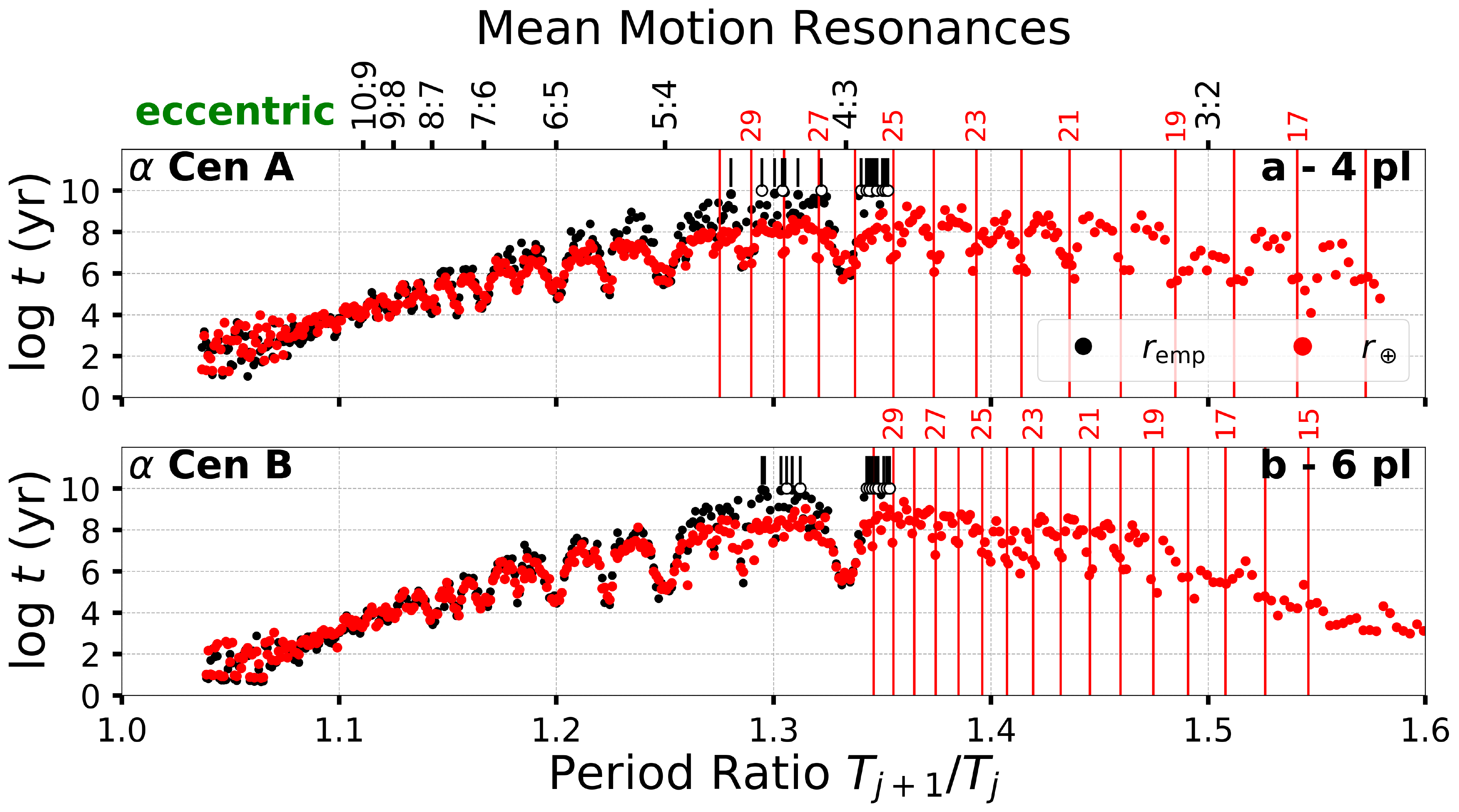}
\caption{Data shown in Figure \ref{46_ecc}, but the horizontal axis has been re-scaled to measure the planetary spacing in terms of the ratio of the initial orbital periods of adjacent pairs of planets.  The locations of first-order mean motion resonances of adjacent planets are marked in black along the top axis. The vertical red lines show the locations of the $N:1$ mean motion resonances of the outermost planet with the binary orbit for the $r_\oplus$ runs, for which they are stronger at a given value of $\beta$,  with odd values of $N$ corresponding to these resonances shown above the plots.  See caption to Figure \ref{235_A} for an explanation of the symbols used.\label{46_res_ecc}}
\end{figure*}

The buildup of interplanetary perturbations near mean motion resonances can destabilize planetary systems, and putative planets in orbit about $\alpha$ Cen A or $\alpha$ Cen B would also experience kicks from the stellar companion.  We focus on the lifetimes of systems around $\alpha$ Cen A as a function of the spacing in period ratios of adjacent planets.  We find similar results for our runs around $\alpha$ Cen B, but because our prescription starts the innermost planet at the inner edge of the host star's HZ, which lies much closer to $\alpha$ Cen B than for $\alpha$ Cen A, strong effects due to the stellar companion only occur at much larger values of $\beta$ for a given planet multiplicity.   Figures \ref{235_res}, \ref{235_res_ecc}, and \ref{46_res_ecc} display local instabilities near mean motion resonances. The contrast is much stronger in systems with minimized free eccentricity; for instance, the instability near the 6:5 resonance appears much sharper in Figure \ref{235_res_ecc}b and c compared to Figures \ref{235_res}b and c.  Systems with minimized free eccentricity are typically longer-lived for a given value of $\beta$ and planetary multiplicity, allowing more time for resonant perturbations to build up.

Figures \ref{235_res}b and \ref{235_res}c show small dips in stability occurring exterior to the 7:5 resonance between adjacent planets.  These dips are due to the high-order $N$:1 mean motion resonances ($N \geq 15$) of the outer planet with the stellar companion. See \cite{Quarles2016} for a discussion of the effecs of these  $N$:1 resonances on the stability of single planets in the $\alpha$ Cen system.  This effect is strongest when the planet begins beyond $\sim$2 AU and depends sensitively on the starting semimajor axis of the planet.  Figure \ref{46_res_ecc} dramatically shows how the other binary component destabilizes planetary systems in which the outermost planet orbits near one of the high-order $N:1$ mean motion resonances with the stellar orbit.

\section{Discussion}
\subsection{Effect of Free Eccentricity}
Starting planets with their forced values of eccentricity significantly improves how tightly planets can be packed on nested prograde orbits in the HZs of $\alpha$ Cen A and  $\alpha$ Cen B.  The maximum value of forced eccentricity that we impart for a single planet is $\sim$0.06, which occurs when the outer planet begins near 2 AU from the host star.  However, the most tightly packed stable systems typically occur where the outermost planet begins with a semimajor axis less than 1.5 AU, and for these configurations our method for forced eccentricity assigns nearly the same value ($e_j = 0.02 - 0.03$) to each of the planets.  

{Our method chooses the initial eccentricity of each planet, $e_j$, individually as a function of its semimajor axis, $a_j$. We find that this methodology usually produces systems with slightly longer lifetimes than setting the initial eccentricity of all of the planets equal to the forced eccentricity of the outermost planet (Fig. \ref{ecsame}).  This is likely stems from starting each planet near its secular equilibrium solution, which keeps the planetary orbits aligned with the binary orbit \citep{Andrade2016,Andrade2017}.}

\subsection{Power Law Fits to Lifetimes of Closely-Packed Systems}
Previous studies of planet packing in simulations with four or more massive bodies (planets + star) have sought to fit the logarithm of the lifetimes of the systems to a linear function of the form: 

\begin{align} \label{eqn:oldfit}
\log\;t = b\beta + c
\end{align}

\noindent in order to evaluate the slope towards stability relative to the spacing parameter $\beta$ \citep{Chambers1996,Smith2009,Obertas2016}. {\bf Their results are summarized in Table \ref{tab:SS_fit}.  We apply this type of analysis independently to each of the 28 sets of our runs whose instability times are plotted in Figs. \ref{235_A}--\ref{235_B_ecc} and \ref{46_ecc}.  We compare our results for differing sets of system parameters to one another and to the single star studies by \cite{Smith2009} and \cite{Obertas2016}, because the types of planetary systems that they simulated are similar to those being evaluated in our study. \cite{Chambers1996} used the same methodology, but because of their more limited computer resources they did not fit over the same range in $\beta$ and they also used different planetary masses.   } 

\begin{deluxetable}{c|cccccc}
\tablecolumns{7}
\tablewidth{0pc}
\tablecaption{Summary Table of Fitted Coefficients from Previous Works\label{tab:SS_fit}}
\tablehead{ \colhead{Ref.} &\colhead{$n_{pl}$} & \colhead{$b$} & \colhead{$\sigma_b$} & \colhead{$c$} & \colhead{$\sigma_c$}& \colhead{$f(2\sqrt{3})$} }
\startdata
\cite{Chambers1996} & 3 & 1.176 & 0.051 & -1.663 & 0.274 & 2.411    \\
\cite{Smith2009} & 3 & 1.496 & -- & -3.142 & -- & 2.040 \\
\hline
\cite{Chambers1996} & 5 & 0.765 & 0.03 & -0.030 & 0.192 & 2.620  \\
\cite{Smith2009} & 5  & 1.012  & -- & -1.686 & -- & 1.820  \\
\cite{Obertas2016} & 5  & 0.951 &-- & -1.202 & -- & 2.092  
\enddata
\tablecomments{Summary of coefficients ($b$ \& $c$) for linear fits to Eq.~(\ref{eqn:oldfit}) from previous studies around single stars with respect to the initial number of planets ($n_{pl}$) in each system.  The range in planetary spacing considered for these fits is $2\sqrt{3} < \beta \leq 8.3$.  We also provide the value of $\log\;t$ when $\beta = 2\sqrt{3}$ (i.e., $f(2\sqrt{3})$).  The dash (--) symbol denotes when uncertainties are not available in the respective work. Uncertainties are not available for any of the previous estimates of $f(2\sqrt{3})$ because the values of the parameters $b$ and $c$ are correlated.}
\end{deluxetable}

Table \ref{tab:lin_fit} provides a summary of our results in terms of the key parameters that describe the linear trends at small $\beta$ for each of our sets of simulations.  The mass of $\alpha$ Cen B is less than that of $\alpha$ Cen A, but the luminosity of $\alpha$ Cen B is substantially lower, which shrinks the width of the associated HZ and shifts {the inner edge of each HZ} closer to $\alpha$ Cen B.  This results in a much steeper slope and indicates that planets orbiting in the HZ of $\alpha$ Cen B are less perturbed by the long-range gravitational effects of $\alpha$ Cen A.  
{Note that our results should \textit{not} be used to extrapolate beyond the fitted region because the binary companion becomes increasingly important as $\beta$ increases, so the systems deviate from the power law dependence found for planetary systems orbiting a single star.  A more sophisticated fitting function is necessary to capture system lifetimes at high $\beta$ values ($\beta \gtrsim 9$).}

{In addition to performing fits analogous to those of previous studies, we also fit our data to a version of Eq.~(\ref{eqn:oldfit}) shifted to place the origin where the separation between planetary orbits is equal to the critical value for two planets orbiting an isolated star:}

\begin{align} \label{eqn:newfit}
\log\;t = b^\prime\beta_{2\sqrt3} + c^\prime,
\end{align}

\noindent  {where $\beta_{2\sqrt{3}} \equiv \beta - 2\sqrt{3}$ and the fitted coefficients in the shifted coordinate system are $b^\prime$ and $c^\prime$. In order to produce Table \ref{tab:lin_fit}, we used the python module {\it{emcee}} \citep{Foreman2013} to determine the best-fitting parameters and their associated uncertainties $\sigma$.  For most of our fits, we include the approximately linear regime, $2\sqrt{3}\leq \beta \leq 8.3$ ($0\leq\beta_{2\sqrt{3}} \leq 8.3 - 2\sqrt{3}\approx 4.836$), in order to compare as fairly as possible with other works.  However for the two planet runs, the smallest value of $\beta$ that produces a stable configuration, $\beta^\dagger$, is typically less than 8.3, and when this is the case we use $\beta^\dagger$ as the outer limit for the fit instead. Note that in all cases $b^\prime \approx b$ and $c^\prime \approx f(2\sqrt{3}) \equiv 2\sqrt3b + c$. }

\begin{figure*}
\epsscale{0.78}
\plotone{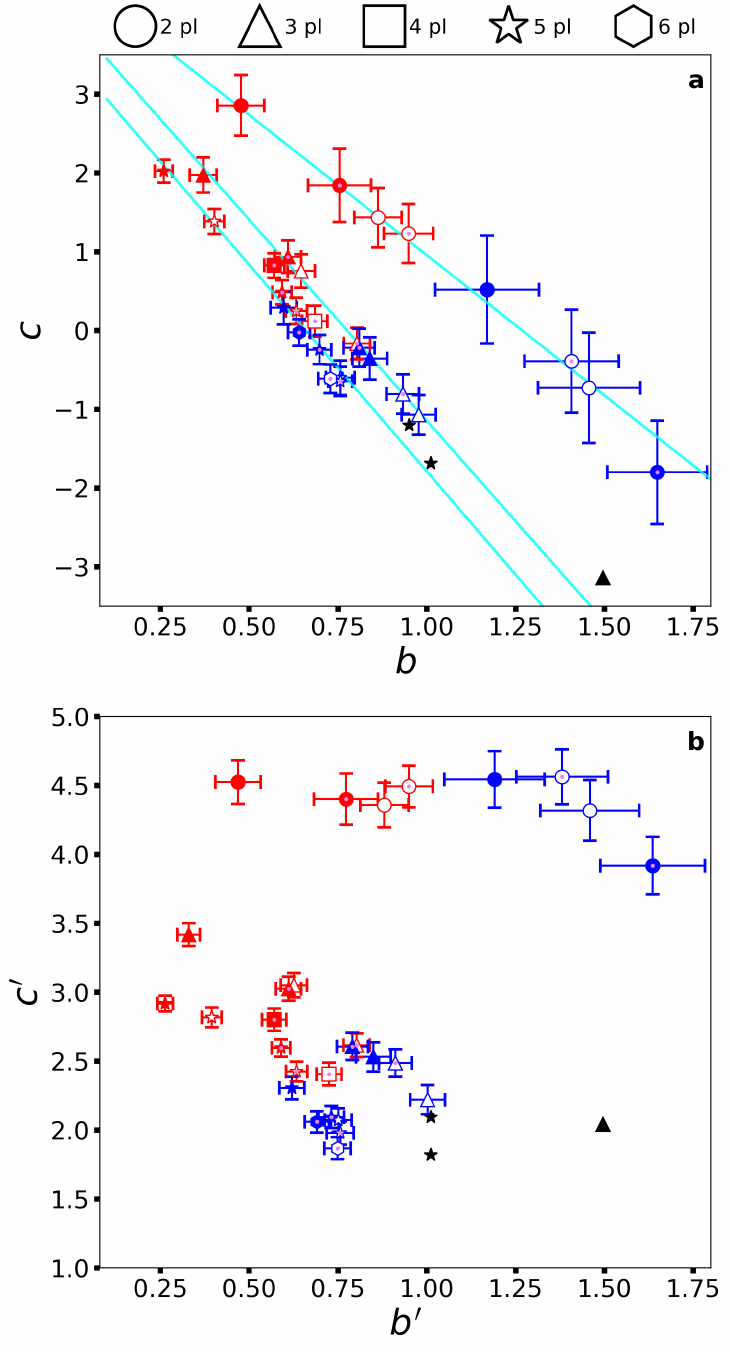}
\caption{{Comparison of the slopes and intercepts of the logarithmic fits given in Equations \ref{eqn:oldfit} and \ref{eqn:newfit}.  Symbol shapes show the number of planets in the system, using the code shown above the plots.  Planetary systems orbiting $\alpha$ Cen A are shown in red, systems orbiting $\alpha$ Cen B use blue, and the black points are for single-star planetary systems \citep{Smith2009,Obertas2016}.  Open and filled symbols correspond to systems that begin at the inner edge of the empirical ($r_{emp}$) or conservative HZ ($r_\oplus$), respectively.  Small pink dots are used to denote systems with initially eccentric planetary orbits.  The numerical values of the points in panels a and b are given in Tables \ref{eqn:oldfit} and \ref{eqn:newfit}, respectively.  Panel a shows the traditional fitting for the coefficients $b$ and $c$, where the cyan lines in this panel show the best linear fit to our results for 2, 3, and 5 planet systems.  Panel b shows fits for the slope ($b^\prime$) and $y$-intercept ($c^\prime$) with $\beta$ shifted by $2\sqrt3$.}  \label{bc_plane}}
\end{figure*}

{Figure \ref{bc_plane}a displays the results of the fits for the coefficients $b$ and $c$ for the various sets of planetary systems that we modeled herein as well as those studied by \cite{Smith2009} and \cite{Obertas2016}.  A strong anti-correlation between $b$ and $c$ is apparent, especially if we restrict our attention to systems with a given number of planets. This trend, combined with the similarity between the lifetimes of systems with planetary separations close to  $ \beta = 2\sqrt{3}$, motivated us to introduce $b^\prime$ and $c^\prime$.  The differences in the slope of the anti-correlation is not much different between three and five planet systems analogous to the limiting effects on stability due to planet multiplicity around single stars \citep{Chambers1996}.  On the other hand, planet multiplicity in $\alpha$ Cen is limited by the outer boundary of stability for single planet systems in binaries \citep{Holman1999}.} 

{Figure \ref{bc_plane}b shows an analogous plot that displays $b^\prime$ and $c^\prime$ for the various sets of planetary systems that we modeled herein, where $b$ and $f(2\sqrt{3})$ are used for the fits in previous studies \citep{Smith2009,Obertas2016} that did not fit their data for $b^\prime$ and $c^\prime$.  The range in the value of $c^\prime$ in Fig.~\ref{bc_plane}b is much less than the range in the value of $c$ in Fig.~\ref{bc_plane}a, with the two planet systems all having very similar values of $c^\prime$.  Transforming to $c^\prime$ demonstrates the similarity in lifetimes of systems with a given number of planets at $\beta={2\sqrt{3}}$ (see Figs. \ref{235_A}--\ref{235_B_ecc}), whereas $c$ estimates system lifetimes substantially outside the fitted region (i.e., $\beta=0$).  Systems with 3--6 planets typically have smaller values of $b^\prime$ and $c^\prime$ than for the two planet systems, and still have an anti-correlation between the parameters plotted, albeit a much weaker one than is present in Fig.~\ref{bc_plane}a. }  

{Considering two, three and five planet systems separately, we fit the (anti-)correlation between $b$ and $c$ and show our result as the three cyan lines in Fig.~\ref{bc_plane}a.  We use the slopes of these lines to determine the shift in $\beta$ required to completely remove the correlation between  $b$ and $c$.  We refer to this shift, whose value is $3.56^{+0.42}_{-0.34}$ for two planet systems, $5.10^{+0.43}_{-0.42}$ for three planet systems and $5.24^{+0.36}_{-0.34}$ for five planet systems, as $\xi$.  The traditional logarithmic fits of systems with the same number planets orbiting different stars or with different initial eccentricities have similar lifetimes near $\beta = \xi$, which implies that the local interplanetary interactions dominate over perturbations from the stellar companion for $\beta \lesssim \xi$.  }

{Figures \ref{bc_plane}a and \ref{bc_plane}b both include eight points representing our results for 2, 3, and 5 planet systems.  These represent a parameter space in which the star being orbited, assumed initial planetary eccentricities, and orbital distance of the inner planet each can take two values. For a given number of planets, single star systems have the steepest slopes.  The initially circular systems orbiting around $\alpha$ Cen A with innermost planet at $r_\oplus$ (filled red points lacking pink dots) have the flattest slopes because they are subjected to the largest eccentricity forcing by the companion star.  In general, for the same number of planets, the less perturbed systems (open symbols with pink dots) tend to have a steeper slope than more perturbed systems, such as those on initially circular orbits or located at larger semimajor axes.  With other parameters held fixed, systems with more planets have shallower slopes.  For very closely spaced planets (small values of $\beta$), perturbations from the companion star are not significant because of shorter system lifetimes and the smaller semimajor axis of the outermost planet.  The lifetimes of such closely spaced systems vary more closely with the orbital periods of the inner planets (Table \ref{tab:HZ})  than with the strength of forcing by the stellar companion.}

\begin{deluxetable}{cc|c|c|c|c|c|c||c|c|c|c|c}
\tablecolumns{13}
\tablewidth{0pc}
\tablecaption{Summary Table of Fitted Coefficients \label{tab:lin_fit}}
\tablehead{ \colhead{Star} & \colhead{$e_o$} & \colhead{$a_1$} & \colhead{$n_{pl}$} & \colhead{$b$} & \colhead{$\sigma_{b}$} & \colhead{$c$} & \colhead{$\sigma_{c}$} & \colhead{$b^\prime$} & \colhead{$\sigma_{b^\prime}$} & \colhead{$c^\prime$} & \colhead{$\sigma_{c^\prime}$} & \colhead{$f(2\sqrt{3})$}}
\startdata
{\multirow{14}{*}{\rotatebox[origin=c]{90}{$\alpha$ Cen A}}} & {\multirow{6}{*}{\rotatebox[origin=c]{90}{circular}}} & 
{\multirow{3}{*}{$r_{\rm emp}$}}& 2 & 0.845 & 0.066 & 1.549 & 0.379 &	0.854 & 0.065 & 4.442	&	0.152 & 4.476	\\
&&& 3 & 0.646 & 0.040 & 0.754 & 0.229 &	0.626 & 0.038 & 3.062	&	0.091	& 2.992\\
&&& 5 & 0.379 & 0.028 & 1.560 & 0.160 &	0.385 & 0.029 & 2.840	&	0.073	& 2.873\\
\cline{3-13}
&& {\multirow{3}{*}{$r_\oplus$}} & 2 & 0.484 & 0.066 & 2.822 & 0.388 & 0.487 & 0.064 & 4.483	&	0.162	& 4.499\\
&&& 3 & 0.351 & 0.038 & 2.121 & 0.217  & 0.359 & 0.031 &	3.331	&	0.079	& 3.337\\
&&& 5 & 0.274 & 0.023 & 1.938 & 0.134  &	0.258 & 0.020 & 2.950	&	0.046	& 2.887\\
\cline{2-13}
& {\multirow{8}{*}{\rotatebox[origin=c]{90}{eccentric}}} & 
{\multirow{3}{*}{$r_{\rm emp}$}} & 2 & 0.953 & 0.072 & 1.212 & 0.392 & 0.952 & 0.073 &	4.527	&	0.162	& 4.513\\
&&& 3 & 0.823 & 0.040 & -0.306 & 0.221 &	0.810 & 0.036 & 2.607	&	0.081	& 2.545\\
&&& 4 & 0.723 & 0.034 & -0.079 & 0.196 & 0.705 & 0.034 &	2.456	&	0.079 & 2.426	\\
&&& 5 & 0.629 & 0.030 & 0.244 & 0.165 &	0.625 & 0.027 & 2.446	&	0.067	& 2.423 \\
\cline{3-13}
&& {\multirow{3}{*}{$r_\oplus$}} & 2 & 0.799 & 0.092 & 1.599 & 0.489 & 0.793 & 0.091 &	4.373	&	0.188 & 4.367	\\
&&& 3 & 0.623 & 0.034 & 0.872 & 0.192  & 0.627 & 0.036 &	2.997	&	0.088 & 3.030	\\
&&& 4 & 0.588 & 0.038 & 0.712 & 0.210 &	0.558 & 0.032 & 2.879	&	0.077 & 2.749	\\
&&& 5 & 0.590 & 0.028 & 0.496 & 0.161  &	0.571 & 0.028 & 2.588	&	0.067	& 2.540\\
\hline
{\multirow{14}{*}{\rotatebox[origin=c]{90}{$\alpha$ Cen B}}} & {\multirow{6}{*}{\rotatebox[origin=c]{90}{circular}}} & 
{\multirow{3}{*}{$r_{\rm emp}$}}& 2 & 1.494 & 0.144 & -0.920 & 0.694  & 1.480 & 0.145 &	4.294	&	0.228	& 4.255\\
&&& 3 & 1.017 & 0.050 & -1.345 & 0.267  & 0.996 & 0.049 &	2.234	&	0.102	& 2.178\\
&&& 5 & 0.758 & 0.039 & -0.587 & 0.211  &	0.742 & 0.038 & 2.084	&	0.087 & 2.039	\\
\cline{3-13}
&& {\multirow{3}{*}{$r_\oplus$}} & 2 & 1.227 & 0.148 & 0.259 & 0.695 & 1.249 & 0.143 &	4.471	&	0.209 & 4.509	\\
&&& 3 & 0.825 & 0.045 & -0.249 & 0.244 & 0.833 & 0.048 & 2.581	&	0.109 & 2.609	\\
&&& 5 & 0.616 & 0.039 & 0.193 & 0.215  & 0.611 & 0.041 & 2.325	&	0.094	& 2.327 \\
\cline{2-13}
& {\multirow{8}{*}{\rotatebox[origin=c]{90}{eccentric}}} & 
{\multirow{4}{*}{$r_{\rm emp}$}} & 2 & 1.371 & 0.136 & -0.177 & 0.655  &	1.417 & 0.133 & 4.499	&	0.210 & 4.572	\\
&&& 3 & 0.931 & 0.044 & -0.789 & 0.236  &	0.919 & 0.042 & 2.474	&	0.092	& 2.436 \\
&&& 5 & 0.744 & 0.037 & -0.575 & 0.204  &	0.792 & 0.038 & 1.891	&	0.083	& 2.002 \\
&&& 6 & 0.764 & 0.031 & -0.795 & 0.173  &	0.737 & 0.034 & 1.888	&	0.079 & 1.852 \\
\cline{3-13}
&& {\multirow{4}{*}{$r_\oplus$}} & 2 & 1.653 & 0.151 & -1.822 & 0.695  & 1.653 &	0.150 & 3.915	&	0.209	& 3.904\\
&&& 3 & 0.823 & 0.043 & -0.343 & 0.235 &	0.796 & 0.045 & 2.593	&	0.101	& 2.508 \\
&&& 5 & 0.679 & 0.034 & -0.116 & 0.191  &	0.707 & 0.037 & 2.146	&	0.089	& 2.236\\
&&& 6 & 0.706 & 0.032 & -0.447 & 0.179 &	0.697 & 0.033 & 2.063	&	0.074	& 1.999
\enddata
\tablecomments{{Summary of coefficients ($b$ \& $c$) and uncertainties ($\sigma_b$ \& $\sigma_c$) for linear fits to $\log\;t = f(\beta) = b\beta + c$ for each of our runs, where these values depend on the host star, initial eccentricity ($e_o$), innermost semimajor axis ($a_1$), and initial number of planets ($n_{pl}$) in the system.  A second set of coefficients ($b^\prime$ \& $c^\prime$) and uncertainties ($\sigma_{b^\prime}$ \& $\sigma_{c^\prime}$) are given where a translation ($\beta_{2\sqrt{3}} = \beta - 2\sqrt{3}$) is applied.  We also provide the value of $\log\;t$ when $\beta = 2\sqrt{3}$ (i.e., $f(2\sqrt{3})$).  The range in $\beta$ considered for these fits are values greater than $2\sqrt{3}$ and include up to the lesser of 8.3 and $\beta^\dagger$.}}
\end{deluxetable}

\begin{deluxetable}{cc|c|c|c|c|c}
\tablecolumns{7}
\tablewidth{0pc}
\tablecaption{Summary of Parameters for the Most Tightly Packed Systems \label{tab:summary}}
\tablehead{ \colhead{Star} & \colhead{$e_o$} & \colhead{$a_1$} & \colhead{$n_{pl}$} & \colhead{$\beta^\dagger$} & \colhead{$n_{HZ}$} & \colhead{$\beta^\ddag$}}
\startdata
{\multirow{14}{*}{\rotatebox[origin=c]{90}{$\alpha$ Cen A}}} & {\multirow{6}{*}{\rotatebox[origin=c]{90}{circular}}} & 
{\multirow{3}{*}{$r_{\rm emp}$}}& 2 & 9.55 & 2 & 12.10 \\
&&& 3 & 22.80 & 3 & 29.30 \\
&&& 5 & -- & -- & -- \\
\cline{3-7}
&& {\multirow{3}{*}{$r_\oplus$}} & 2 & 21.30 & 2 & 24.50 \\
&&& 3 & -- & -- & -- \\
&&& 5 & -- & -- & -- \\
\cline{2-7}
& {\multirow{8}{*}{\rotatebox[origin=c]{90}{eccentric}}} & 
{\multirow{4}{*}{$r_{\rm emp}$}} & 2 & 7.95 & 2 & 9.75 \\
&&& 3 & 10.75 & 3 & 13.15 \\
&&& 4 & 13.60 & 4 & 16.60\\
&&& 5 & -- & -- & -- \\
\cline{3-7}
&& {\multirow{4}{*}{$r_\oplus$}} & 2 & 7.45 & 2 & 8.60 \\
&&& 3 & 12.90 & 3 & 17.15 \\
&&& 4 & -- & -- & -- \\
&&& 5 & -- & -- & -- \\
\hline
{\multirow{14}{*}{\rotatebox[origin=c]{90}{$\alpha$ Cen B}}} & {\multirow{6}{*}{\rotatebox[origin=c]{90}{circular}}} & 
{\multirow{3}{*}{$r_{\rm emp}$}}& 2 & 6.75 & 2 & 8.60 \\
&&& 3 & 9.95 & 3 & 11.90 \\
&&& 5 & 22.40 & {\bf{4}} & -- \\
\cline{3-7}
&& {\multirow{3}{*}{$r_\oplus$}} & 2 & 6.30 & 2 & 7.35 \\
&&& 3 & 18.30 & 3 & 21.50 \\
&&& 5 & -- & -- & -- \\
\cline{2-7}
& {\multirow{8}{*}{\rotatebox[origin=c]{90}{eccentric}}} & 
{\multirow{4}{*}{$r_{\rm emp}$}} & 2 & 6.65 & 2 & 8.25 \\
&&& 3 & 8.95 & 3 & 11.25 \\
&&& 5 & 12.65 & 5 & 14.20 \\
&&& 6 & 13.50 & {\bf{5}} & 15.80 \\
\cline{3-7}
&& {\multirow{4}{*}{$r_\oplus$}} & 2 & 6.35 & 2 & 7.25 \\
&&& 3 & 9.90 & 3 & 12.10\\
&&& 5 & 13.90 & {\bf{4}} & 16.20 \\
&&& 6 & -- & -- & --
\enddata
\tablecomments{ Summary of most tightly-packed stable (lifetime $>$ 6 Gyr) value of $\beta$ ($\beta^\dagger$), number of planets in the HZ at $\beta^\dagger$ ($n_{HZ}$), and the 10$^{th}$ value of $\beta$ from our simulations that remained stable until our integrations were terminated at 10 Gyr ($\beta^\ddag$).  These values depend on the host star, initial eccentricity ($e_o$), innermost semimajor axis ($a_1$), and initial number of planets ($n_{pl}$) in the system.  {\bf{Boldface}} values indicate when the initial number of planets in the HZ is less than the initial number of planets in the system.  The dash (--) symbol represents sets of systems that survived for $< 6$ Gyr for all values of $\beta$ that we simulated ($\beta^\dagger$) or for which we did not find 10 configurations that survived up to 10 Gyr ($\beta^\ddag$). }
\end{deluxetable}

\section{Conclusions}

We have numerically integrated tightly-packed systems containing from 2--6 planets, each being one Earth mass (1~M$_\oplus$),  on prograde orbits in (and in some cases slightly exterior to) the habitable zones of our stellar neighbors, $\alpha$ Centauri A and B, for up to 10 Gyr.  The orbits of the planets are initially uniformly spaced in units of the mutual Hill radius of the planets, $R_{H_{j,j+1}}$, the value of which is given by Eq.~(\ref{eqn:hill}).  Table \ref{tab:summary} summarizes our results and lists the smallest difference in initial orbital semimajor axes of neighboring planets in units of $R_{H_{j,j+1}}$ that survives longer than 6 Gyr  (the nominal age of the $\alpha$ Centauri system), $\beta^\dagger$; the number of those planets initially within the designated HZ, $n_{HZ}$; and the tenth value of orbital separation parameter $\beta$ for which the system survived for the entire 10 Gyr simulated, $\beta^\ddag$.

Perturbations from the companion star substantially reduce the stability of closely-spaced multi-planet systems orbiting within the habitable zones of $\alpha$ Centauri A and B relative to that of comparably-spaced planets orbiting single stars.  The most closely-spaced two planet system that we found to survive for at least 6 Gyr was in orbit around $\alpha$ Cen B and initially had $\beta = 6.3$, which is much greater than the minimum value of $\beta = 2\sqrt{3} \approx 3.464$ required for Hill stability around single stars.  The companion star induces a ``forced'' eccentricity on circumstellar planetary orbits within a binary system, and starting planets with orbital eccentricities close to their forced eccentricities generally allowed for more closely-packed systems than for planets with initially circular orbits.

We found that three 1~M$_\oplus$ planets can stably orbit around $\alpha$ Cen A within the conservative HZ and four planets can stably orbit within its optimistic (empirical) HZ.  For $\alpha$ Cen B, four planets can survive in the conservative HZ, while five can remain in its optimistic HZ.   We also found cases where one additional planet could orbit slightly exterior to the HZ of $\alpha$ Cen B, so it is quite possible that with a higher resolution grid in $\beta$ and/or a slightly different choice for initial planetary eccentricities, five planets might survive in the conservative HZ and six could remain in its optimistic HZ. (See Table \ref{tab:outer_HZ} for HZ limits in terms of $\beta$.)  Spacing planets non-uniformly may also allow more planets to orbit stably within these stars' HZs, as could more exotic configurations such as retrograde trajectories relative to the binary orbit and co-orbital (e.g., Trojan) planets, but we leave studies of these configurations to future work. 

Overall, at least nine planets could orbit within habitable zones of the stars $\alpha$ Centauri A and B.  However, more studies are needed to fully understand the possible formation environment for these types of systems and which types of initial architectures would be more typical.  Our results demonstrate some of the various dynamical effects possible within the lifetime of the binary system. If all of the planets initially orbit well interior to the single-planet stability limit found by \cite{Quarles2016}, then perturbations from the companion star have only a slight destabilizing effect on three and five planet systems; however, they are still fundamentally important to two planet systems.  The regime near the outer stability limit exhibits strong destabilizing perturbations of the outermost planet in the system by long-term external forcing due to $N:1$ resonances with the stars' mutual orbit.   Fortuitously, multi-planet systems can survive in large regions of the HZs of both stars, and these regions should be the focus of efforts to detect many potential worlds orbiting around our closest stellar neighbors.

\acknowledgments
{The authors thank the anonymous referee for a careful and thoughtful review that resulted in improving the clarity and quality of the manuscript.}  B. Q. gratefully acknowledges support by an appointment to the NASA Postdoctoral Program at the Ames Research Center, administered by Oak Ridge Associated Universities through a contract with NASA.  Some of this project was performed at the OU Supercomputing Center for Education \& Research (OSCER) at the University of Oklahoma (OU).  We thank R. Belikov and N. Kaib for helpful comments on the manuscript.

\bibliographystyle{apj}
\bibliography{bibliography}

\end{document}